\documentclass[12pt,preprint]{aastex}

\usepackage{graphicx}
\usepackage{rotating}
\shorttitle{}
\shortauthors{Nesvorn'y et al.}

\begin{document}
\baselineskip 19.pt

\title{Discovery of 63 New Young Asteroid Families}

\author{David Nesvorn\'y$^1$, David Vokrouhlick\'y$^2$, Miroslav Bro\v{z}$^2$, Fernando V. Roig$^3$  }

\affil{(1) Solar System Science \& Exploration Division, Southwest Research Institute, 1301 Walnut St., 
  Suite 400,  Boulder, CO 80302, USA}

\affil{(2) Institute of Astronomy, Charles University, V Hole\v{s}ovi\v{c}k\'ach 2, 
  CZ–18000 Prague 8, Czech Republic}

\affil{(3) Observat\'orio Nacional, Rua Gal. Jose Cristino 77, Rio de Janeiro, RJ 20921-400, Brazil}

\begin{abstract}
  We searched for young asteroid families -- those with ages $t_{\rm age} < 10$ Myr and at least three
  members -- using the proper element catalog from Nesvorn\'{y} et al. (2024a). Our approach employed
  the Hierarchical Clustering Method (HCM) in a five-dimensional space of proper orbital elements:
  semimajor axis, eccentricity, inclination, proper nodal longitude, and proper perihelion longitude.
  The proper longitudes were calculated for various times in the past. Any convergence of these angles
  at times $t < 10$ Myr ago was automatically identified by our algorithm as a clustering event in
  5D space at time $t$. Using this method, we successfully recovered all previously known young families
  (over 40) and discovered 63 additional ones. The formation ages of these families were determined
  through backward orbital integrations. To validate orbital convergence, we applied three different
  methods and obtained generally consistent results. Notably, the vast majority of identified young
  families have the formation ages $t_{\rm age} \lesssim 1$ Myr. The number and properties of these families
  provide valuable constraints on the frequency of recent large cratering or catastrophic collisions,
  offering new insights into the ongoing collisional evolution of the main asteroid belt. Alternatively,
  at least some of the families identified here could have been produced by the spin-up and rotational
  fission of their parent bodies. Future studies should address the relative importance of collisions
  and rotational fission for young asteroid families identified here.   
\end{abstract}

\section{Introduction}

It is difficult to point out a scientific subject that is as fundamentally linked to so many important research 
areas in planetary science as studies of {\it asteroid families} (Hirayama 1918). An asteroid family consists of dispersed 
pieces of a parent body that suffered a large impact (see below for a discussion of rotational fission; Pravec et al.
2018). Here, the telescopic observations of fragments offer a unique opportunity to examine the interior of the parent
body and learn about such cardinal physical processes as the primary accretion, thermal processing, geophysical
differentiation, etc. Boulders released by a family-forming event may, with some delay, end up on near-Earth orbits
and fall as meteorites (Wisdom 1985, Marsset et al. 2024, Bro\v{z} et al. 2024a,b). 

Studies of asteroid families help us to understand the physics of large scale collisions, a process by
which the Earth and other terrestrial planets formed. In the main belt, where dozens of asteroid families were
identified (Novakovi\'c et al. 2022), they provide key constraints on the collisional evolution of asteroids, with some
works suggesting that practically the whole belt may be the result of early, unresolved breakups 
(Bottke et al. 2005, Delbo et al. 2017, Dermott et al. 2018). The asteroid families are also instrumental 
to our understanding of the orbital evolution of asteroids, including the radiation effects (Yarkovsky and 
YORP; Vokrouhlick\'y et al. 2015) and resonant interactions -- processes that underpin the dynamical origin 
of near-Earth asteroids and meteorites (Wisdom 1985, Vokrouhlick\'y \& Farinella 2000).

The detection of asteroid families with young formation ages, $t_{\rm age} \lesssim 10$ Myr, is one of the highlights 
of asteroid research. A poster child of this exciting development is the Karin family, part of the larger Koronis family,
that was shown to have formed $5.8 \pm 0.2$ Myr ago (Nesvorn\'y et al. 2002). The Karin family was identified by the
traditional means,  using the Hierarchical Clustering Method (HCM; Zappal\`a et al. 1990) on {\it proper} orbital
elements (Section 2).\footnote{The Karin family had 39 members back in 2002 when the original work was published.
It now has over 2000 members (Table 1).}  
The age of the Karin family was established by numerically integrating the orbits of member
asteroids back in time to show their past convergence. There are now $\simeq 43$ known young families with formation
ages between 15 kyr and $\sim 15$ Myr (Table 1; Nesvorn\'y et al. 2015, 2024a; Carruba et al.
2018a; Pravec et al. 2018, 2019a; Novakovi\'c \& Radovi\'c 2019; Fatka et al. 2020; Vokrouhlick\'y et al. 2024a).

{\it Significance of young asteroid families.}
The young families are important because collisional and dynamical processes had little time to act on these families
to alter their properties. The young families therefore attract much attention from researchers studying impact
physics, space weathering, debris disks, etc. Specifically: {\bf (i)} The radiation forces do not have enough time
to modify the orbital distribution of fragments produced by recent breakups. The young families can therefore
be used to probe the physics of large scale collisions (Michel et al. 2015). {\bf (ii)}
The recent breakups are sources of the zodiacal dust bands (Sykes \& Greenberg 1986, Nesvorn\'y et al. 2006a, Marsset
et al. 2024). By studying
them we can learn things relevant to the origin of debris disks (Wyatt 2008). {\bf (iii)} Tracers of the Veritas
family breakup can be found in the measurements of extraterrestrial $^3$He in $\simeq$8.2 Myr old Earth sediments
(Farley et al. 2006). This opens a whole new interdisciplinary research area that links the terrestrial accretion
record to astronomical events. {\bf (iv)} The surfaces of asteroids in the recently-formed families are geologically
young. Their spectroscopic properties are the point of departure for space weathering processes (Jedicke et al. 2004,
Vernazza et al. 2009). {\bf (v)} Several Main Belt Comets (MBCs) are members of young families. This relationship
can help us to understand how MBCs become activated (Hsieh et al. 2018).

It is thought that some (small) asteroid families could have been produced by rotational fission of a parent body
when two or more fragments became unbound (Pravec et al. 2018, Fatka et al. 2020). A good example of this is the
Lucascavin family with only three known members (Vokrouhlick\'y et al. 2024a). Rotational fission is thought to
be the main source of asteroid pairs (Vokrouhlick\'y \& Nesvorn\'y 2008; Pravec et al. 2010, 2019a). Pravec et al. (2010)
pointed out that asteroid pairs show a correlation between the rotation period of the primary (i.e., the larger body in a
pair), $P_1$, and the absolute magnitude difference between the primary and secondary (i.e., the second largest body
in a pair), $\Delta H$. The correlation is consistent with the transient binary formation by rotational fission of the
parent body, and the subsequent requirement for secondary's escape (Pravec et al. 2010). Pravec et al. (2018) extended
the rotational fission model from pairs to young/small asteroid families. They argued that the majority of young/small
families studied by them (11 out of 13) show the same trend of $P_1$ vs. $\Delta H$ as asteroid pairs, which could be
an indication that these young/small families formed by rotational fission. We discuss the relative importance 
of collisions and rotational fission in Section 3.6.
  
In this work we searched for young asteroid families in the osculating and proper orbital element catalogs that have roughly
tripled in size since our last systematic effort in this direction (Nesvorn\'y et al. 2015). In total, we identified
63 new cases which brings the total of young asteroid families, ages $t_{\rm age}<10$ Myr, to over a hundred.  We
determined the formation ages of young families by backward integrations, constrained their formation conditions and,
at least a few cases, inferred the possible drift of individual family members by radiation effects (e.g., Nesvorn\'y
\& Bottke 2004, Carruba et al. 2016). The new catalog of young families (Tables 1--3) can be used to constrain the
collisional evolution of main belt asteroids. We start by describing the methods in Section 2, and proceed by reporting
the results in Section 3. Conclusions are given in Section 4. 

\section{Methods}

\subsection{Osculating elements}

Shortly after an impact, the fragments (and reaccumulated bodies) launched from a parent body will separate 
from each other. Initially, they will have similar orbits with nearly the same values of osculating orbital angles:
the nodal longitude $\Omega$, perihelion longitude $\varpi$ and mean longitude $\lambda$. The orbits will subsequently
diverge due to the (i) {\it Keplerian shear} from slightly different orbital periods, and (ii) {\it differential
  precession} driven by planetary perturbations. As for (i), the dispersal of fragments along the orbit is relatively
fast, and the clustering in $\lambda$ is not expected if a family is older than $\sim$100-10,000 yr. As for (ii),
$\Omega$ and $\varpi$ diverge on a time scale $T_f = \pi/(a\; \partial f / \partial a) (V_{\rm orb}/\delta V)$, where $f$
is either the nodal precession frequency $s$ or the apsidal precession frequency $g$, $V_{\rm orb}$ is the orbital speed,
and $\delta V$ is the ejection speed.

For example, $\partial s / \partial a \simeq -70$
arcsec yr$^{-1}$ au$^{-1}$ and $\partial g / \partial a \simeq 94$ arcsec yr$^{-1}$ 
au$^{-1}$ for the Karin family ($a \simeq 2.865$~AU; Nesvorn\'y et al. 2002). With $\delta V=15$ m s$^{-1}$ (Nesvorn\'y 
et al. 2006b) and $V_{\rm orb}=17.7$ km s$^{-1}$, this gives $T_s=3.8$ Myr and $T_g=2.8$ Myr. Since $t_{\rm age} > T_s$ 
and $t_{\rm age} > T_g$ in this case, $\Omega$ and $\varpi$ of family members are not expected
to be clustered at the present time (indeed they are {\it not} clustered).
Conversely, the clustering of $\Omega$ and $\varpi$ would be expected for families with 
$t_{\rm age} \lesssim 1$ Myr. This expectation leads to the possibility that the very young collisional families  
could be detected in the catalogs of {\it osculating} orbital elements, where they should show up as clusters in 
5D space of $a$, $e$, $i$, $\varpi$ and $\Omega$. This method was first successfully applied in practice to the 
Datura family ($t_{\rm age} \simeq 0.5$ Myr; Nesvorn\'y et al. 2006c, Vokrouhlick\'y et al. 2017). 

The osculating orbital elements are subject to short-periodic oscillations (periods comparable to the orbital 
period). As these oscillations evolve out of phase for different family members, the initially tight concentration 
of orbits becomes dispersed. It is therefore useful, at least in some cases, to use the {\it mean} orbital 
elements (Ro\.{z}ek et al. 2011), with the short-periodic oscillations being removed by a low-pass filter, 
or even the proper elements, with the proper angles $\Omega_{\rm p} = s t + \phi_\Omega$ and $\varpi_{\rm p}
= g t + \phi_\varpi $ being defined from the Fourier analysis (Section 2.2; Nesvorn\'y et al. 2024a, hereafter NRVB24).

\subsection{Proper elements}

Our algorithm for young family identification takes advantage of the proper element catalog published in NRVB24.
NRVB24 selected all orbits of main belt asteroids from the Minor Planet Center (MPC) catalog on February 9, 2024.
The osculating orbits were given at the JD 2460200.5 epoch. The planetary orbits (Mercury to Neptune) were obtained
for the same epoch from the DE 441 Ephemerides (Park et al. 2021). All orbits were numerically integrated with
the \texttt{Swift} integrator (Levison \& Duncan 1994; code \texttt{swift\_{}rmvs4}), which is an efficient
implementation of the Wisdom-Holman map (Wisdom \& Holman 1991). NRVB24 used a short time step (1.1 days) and
integrated all orbits backward in time for 10 Myr. The backward integration is useful to identify any past
convergence of angles, which may indicate the formation time of a young asteroid family (see below).

The Frequency Modified Fourier Transform (FMFT; \v{S}idlichovsk\'y \& Nesvorn\'y 1996, Laskar 1993) was applied
in NRVB24 to obtain a Fourier decomposition of each signal. They used the complex variable $x(t) + \iota y(t)$ with
$x=e\cos(\varpi)$ and $y=e\sin (\varpi)$ for the proper eccentricity, and $x=\sin(i)\cos(\Omega)$ and $y=\sin(i) \sin (\Omega)$
for the proper inclination, where $\varpi$ and $\Omega$ are the perihelion and nodal longitudes. FMFT was first applied to
planetary orbits to obtain the planetary frequencies $g_j$ and $s_j$, governing the perihelion and nodal precession,
respectively. The forced terms with these frequencies were identified in the Fourier decomposition of each asteroid orbit
and subtracted from asteroid's $x(t) + \iota y(t)$. The FMFT was then applied to all asteroid orbits to compute the
frequencies $g$ and $s$, and phases $\phi_\Omega$ and $\phi_\varpi$. These results allow us to compute $\Omega_{\rm p}$
and $\varpi_{\rm p}$ at any time in the past 10 Myr.

\begin{figure}
\epsscale{0.9}
%\hspace*{-1.7cm}\plotone{proper1.pdf}
\plotone{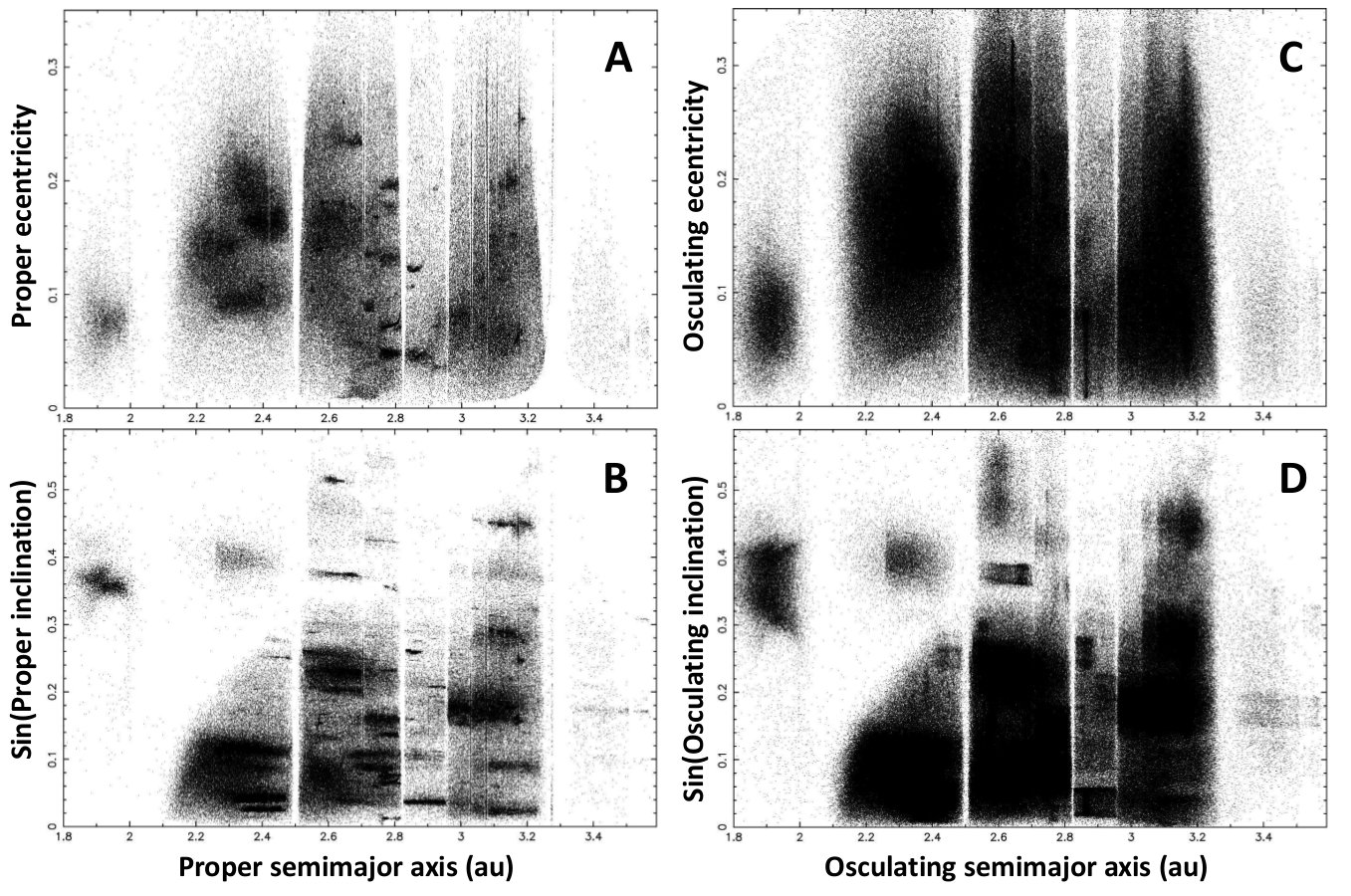}
\caption{Proper (panels A and B) and osculating orbits (panels C and D) of main belt asteroids. 
  In the proper element space, various orbital structures come into focus.}
\label{proper1}
\end{figure}

The proper elements $e_{\rm p}$ and $\sin i_{\rm p}$ were computed in NRVB24 as the mean of $\sqrt{x(t)^2+y(t)^2}$, with the
forced terms removed, over a relatively long interval (5 Myr). Following Kne\v{z}evi\'c \& Milani (2000), the proper
semimajor axis was computed as the mean value of the osculating semimajor axis over the same time interval.
The new catalog of proper elements for 1,249,051 asteroids is available at \texttt{https://asteroids.on.br/appeal/},\\
\texttt{www.boulder.swri.edu/\~{}davidn/Proper24/}, and the PDS node\\
(\texttt{https://sbn.psi.edu/pds/resource/doi/nesvornyfam\_2.0.html}).
See Fig. \ref{proper1} for the illustration of orbital distribution of asteroids in osculating and proper elements. 
  
\subsection{Identification of young asteroid families}

We developed a new method to identify young asteroid families. It consists in applying the
Hierarchical Clustering Method (HCM; Zappal\`a et al. 1990) in five dimensions. The 5D metric was defined as
\begin{equation}
d = {3 \times 10^4\, {\rm m/s} \over \sqrt{a_{\rm p}}} \sqrt{ {5 \over 4} 
  \left({\delta a_{\rm p}/a_{\rm p}}\right)^2 + 2 (\delta e_{\rm p})^2 + 2 (\delta \sin i_{\rm p})^2 +
  k_\Omega  (\delta \Omega_{\rm p}(t))^2 + k_\varpi  (\delta \varpi_{\rm p}(t))^2}\ , 
\label{metric}
\end{equation}
where $3 \times 10^4\, {\rm m/s}$ is the orbital speed at 1 au, $\delta$ indicates differences in the proper elements
between two neighbor orbits, and $k_\Omega=k_\varpi=10^{-7}$ (Nesvorn\'y et al. 2006c). The HCM algorithm
clusters bodies by linking them together in a chain where the length of each segment is required to be
$d < d_{\rm cut}$, with a user-defined cutoff parameter $d_{\rm cut}$.

Our systematic search for young families was conducted by initializing chains from {\it every} asteroid in the NRVB24
catalog. We explored different times in the past and evaluated the metric in Eq. (\ref{metric}) with
$\Omega_{\rm p} = s t + \phi_\Omega$ and $\varpi_{\rm p} = g t + \phi_\varpi$ for a hundred values in the
$0 \leq t \leq 10$ Myr interval (a 0.1 Myr spacing). The very young families with $t_{\rm age} \lesssim 1$ Myr are still expected to
have $\Omega_{\rm p}(t)$ and $\varpi_{\rm p}(t)$ clustered at the present time ($t=0$).
The young asteroid families such as Karin or Veritas are expected to
show 5D clustering at $t \sim t_{\rm age}$. Our method can therefore identify all young families assuming that they
show 5D clustering in the explored time interval.
It is preferable to operate in 5D rather than in 3D, because the increased number of dimensions 
improves our chances to detect a statistically significant cluster with only a few members. Moreover,
it is better to use the proper elements than the osculating elements because the proper elements are not subject
to short-period variations (e.g., Ro\.{z}ek et al. 2011).

We tested several cutoff
distances, $d_{\rm cut}$, and visualized the results with an in-house software package that interactively displays
the distribution of proper orbits in a selected zone, allows the user to zoom out and zoom in, and perform any kind
of active rotation.\footnote{A well defined family is a compact group in the proper element space that stands out from the background.
In such a situation, there is often a range of cutoff distances for which the membership does not change much.
The visualization software helps us to optimize the cutoff distance as we can inspect groups identified
with different cutoffs, make sure that we are not missing some obvious members or grabbing nearby groupings 
that would not make sense physically. It also allows us to check how nearby resonances may be affecting the
family membership at different cutoffs. All these choices would be difficult to make blindly or with an automated algorithm.}
The rotation is particularly useful because the user can easily check whether any concentration
seen in a projection is a real concentration of proper orbits (the statistical significance of new families
is discussed in Section 3.4). For each confirmed family, we identified the lowest numbered asteroid that appeared
to be associated with 
the family and used it to label the family. This association is sometimes not unique as there are two or more large
bodies in/near the family, with some being more or less offset from the family center. Some of the more ambiguous
cases are noted in Tables 2 and 3. The bulk of the families reported in Tables 2 and 3 were identified with
a relatively strict cutoff of $d_{\rm cut}=10$ m s$^{-1}$. The new young families are very well separated from the
background such as this fixed choice of cutoff is appropriate in the majority of cases.

\subsection{Asteroid family ages}

We estimated $t_{\rm age}$ for each new family. This was done by numerically integrating the 
orbits of family members back in time in an attempt to identify their past convergence.
The integrations accounted for: {(i)} the present uncertainty of orbits, {(ii)} the Yarkovsky effect, 
and {(iii)} planetary perturbations. To account for {(i)}, we cloned orbits of each member asteroid
to sample the orbital elements within the uncertainty interval.\footnote{The orbital uncertainties of asteroid orbits
  were obtained from the NASA JPL Horizons system.} 
As for {(ii)}, the clones were assigned
realistic values of ${\rm d}a/{\rm d}t$ inferred from the observational detection of the Yarkovsky effect and
theory (Vokrouhlick\'y et al. 2015).

For example, asteroid (101955) Bennu with the diameter
$D_{\rm Bennu} \simeq 0.5$ km, semimajor axis $a_{\rm Bennu}=1.126$ au and obliquity $\theta \simeq 178^\circ$ has
${\rm d}a/{\rm d}t = (-19.0 \pm 0.1) \times 10^{-4}$ au Myr$^{-1}$ (Chesley et al. 2014, Greenberg et
al. 2020). Thus, for a C-complex family in the main belt, we used clones with
$|{\rm d}a/{\rm d}t| \leq 1.9 \times 10^{-3} (D_{\rm Bennu}/D) (a_{\rm Bennu}/a)^2$ au Myr$^{-1}$.
The maximum drift for an S-complex family was scaled from this value by accounting for higher bulk density
and higher albedo of S-complex asteroids. The highest possible drift rate was used for families with an unknown
taxonomic type. The {\tt swift\_rmvs4} code (Levison \& Duncan 1994) was modified to account for ${\rm d}a/{\rm d}t$ from 
the Yarkovsky effect. The YORP effect was ignored. Inferred values ${\rm d}a/{\rm d}t$ for an individual
body thus stand for the {\it average} drift rate of that body over the family age (Nesvorn\'y \& Bottke 2004,
Carruba et al. 2016).

The backward integrations were run to times $t < 10$ Myr (shorter integrations were used for the very young
families, longer for older ones). The results of backward integrations in NRVB24, which did not account
for effects {(i)} and {(ii)}, were used to choose the appropriate integration times.
Integrations past 10 Myr are not required because the orbital history of asteroids cannot be deterministically 
reconstructed over very long timescales.
The orbital convergence has been established by following the criteria 
developed previously (Nesvorn\'y et al. 2006c, Vokrouhlick\'y \& Nesvorn\'y 2008; see examples in Section 3.3).
The family age was estimated as the time in the past with the strongest orbital convergence. Conservative
uncertainties were assigned in each case (Section 3.3).   

The backward integrations described above require individual approach in each case. It is difficult to systematically
apply this method for the large number of young families identified here. Therefore, to complement this
approach and establish how different approximations may affect the age estimate we also applied two related methods.

\begin{figure}
\epsscale{1.0}
%\plotone{3152_Jones.JPG}
%\plotone{conv_3152.JPG}
%\plotone{conv_10321.JPG}
%\plotone{conv_18429.JPG}
%\plotone{conv_108138.JPG}
%\plotone{convergence1.JPG}
\plotone{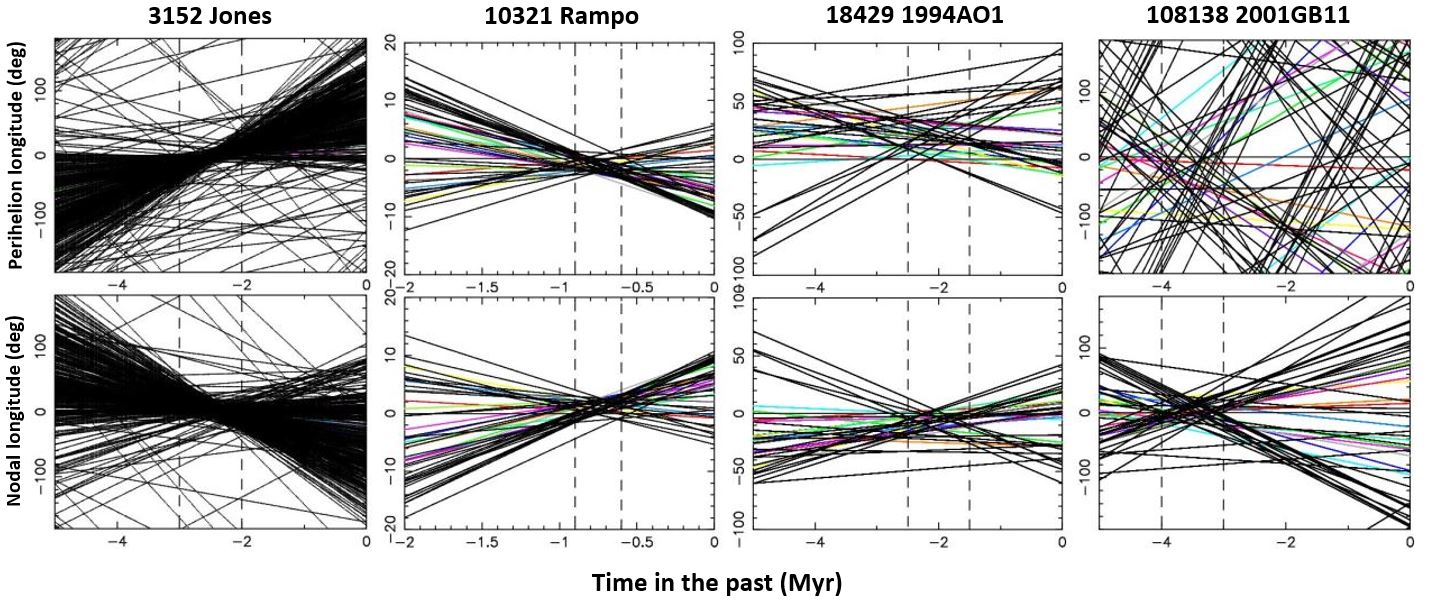}
\caption{The convergence of proper angles for 3152 Jones, 10321 Rampo, 18429 1994AO1 and 108138 2001GB11,
  illustrating the method based on proper angles for previously known young families. All members identified by 3D HCM
  are plotted. A few of the nominal members that do not participate in the convergence may be interlopers.
  See Table 1 for the cutoff distance and number of family members in each case. The vertical dashed lines
  delimit the admissible age interval. 108138 2001GB11 shows a relatively poor
  convergence in the proper perihelion longitude. The age estimate is based on the nodal convergence in
  this case. Note that the 108138 2001GB11 family does not have the orbital angles clustered at the present
  epoch and was identified by our method when the algorithm searched for 5D clustering $\sim 3$-4 Myr in the past. This,
  to a lesser degree, applies to other families shown here as well.}
\label{case1}
\end{figure}

The first method consists in establishing the convergence of {\it proper} angles obtained from NRVB24. The advantage
of this method is that no additional integrations were needed as the proper angles for each asteroid can be
computed from the proper frequencies and proper angles reported in NRVB24 (Sections 2.2 and 2.3). This method is also easy
to automate. The downside is that it ignores the effects (i) and (ii) discussed above, which can be important
for very small, fast drifting members with poorly determined orbits. In addition, the proper elements were determined
from a 5-Myr time span in NRVB24, which may be problematic in cases where there is significant chaotic evolution
of orbits over this interval.\footnote{The use of osculating elements would be favored in this case.} Figure~\ref{case1}
illustrates this method for four previously known young asteroid families:
3152 Jones ($t_{\rm age}=2.5\pm0.5$ Myr), 10321 Rampo ($t_{\rm age}=0.8\pm0.1$ Myr),\footnote{The Rampo family was previously
  estimated to be $0.78^{+0.13}_{-0.09}$ Myr old (Pravec et al. 2018). Based on the convergence tests for Rampo individual
  family members, Pravec et al. (2018) also found possible evidence for a second event $\sim 1.4$ Myr ago.} 18429 1994AO1
($t_{\rm age}=2.0\pm0.5$ Myr) and 108138 2001GB11 ($t_{\rm age}=3.5\pm0.5$ Myr).

The second method consists in backward integration {\it without} the effects (i) and (ii) discussed above,
with the goal of establishing the past orbital convergence from the largest family members. The large members
may have relatively well determined orbits and did not excessively drift by the Yarkovsky effect over the young
family age. This method is similar to the one described above -- for the proper angles -- but here we used the
{\it osculating} angles. The results of the three methods described above were synthesized into the best estimate of each
family's age (Section 3.3). 

\section{Results}

\subsection{Previously reported young families} 

We first collected all young families reported in previous works: they were 43 in total. Table 1 gives the list
of these families together with the relevant references. We examined all these families in detail and determined
the appropriate cutoff distance for each of them in the NRVB24 catalog. Table 1 reports the number of members of
every known young family with the preferred cutoff distance. The great majority of these families are real beyond doubt,
as demonstrated in the original publications. In one case, 15156 2000FK38, we could not establish
that the (alleged) clustering was statistically significant (see NRVB24 and Section 3.4 for tests of
statistical significance). This case would need to be examined in a greater detail. The 15156 2000FK38 family was
proposed in Novakovi\'c et al. (2012) as related to the main belt comet P/2006 VW139. It was listed
as a candidate family in Nesvorn\'y et al. (2015). In addition, there is no clear consensus about the correct
age of several young families. This most notably applies to the Florentina, Beagle and Kutaisii families.
Some of these families could potentially be older than 10 Myr.  

We noted several cases of known (young) families that were not reported in NRVB24 and were missing from the
related catalog.\footnote{The catalog of NRVB24 families is available from \texttt{www.boulder.swri.edu/\~{}davidn/Proper24/}
  and \texttt{https://sbn.psi.edu/pds/resource/doi/nesvornyfam\_2.0.html}.} NRVB24 reported 153 new asteroid families
in the main asteroid belt that were not
listed in Nesvorn\'y et al. (2015). There already were 122 asteroid families listed in Nesvorn\'y et al. (2015): 114
in the main asteroid belt (the Nysa-Polana complex, FIN 405, is counted as three families, Nysa/Mildred, Polana/Eulalia
and New Polana), 6 families in Jupiter Trojans, and 2 families in Hildas in the 3:2 resonance with Jupiter.  One of the
families reported in Nesvorn\'y et al. (2015), (709) Fringilla (FIN 623), was split into two overlapping families in
NRVB24, (19093) 1979MM3 and (37981) 1998HD130. There therefore were 268 known families in the main asteroid belt,
plus 8 known families in the resonant Hilda and Trojan population, reported in Nesvorn\'y et al. (2015, 2024a), for the
total of 276.

The joint catalog published in Nesvorn\'y et al. (2015) and NRVB24 (HCM Asteroid Families V3.0 and HCM Asteroid
Families Bundle V2.0 on the PDS node) was supposed to be the complete census of asteroid families known to date.
In this work, however, we realized that several known families were omitted from NRVB24 and the joint catalog. These cases
are: 525 Adelaide, 2258 Viipuri, 4765 Wasserburg, 5026 Martes, 
5478 Wartburg, 6825 Irvine, 10321 Rampo, 10484 Hecht, 11842 Kap'bos, 18777 Hobson, 22280 Mandragora, 39991 Iochroma,
63440 2001MD30, 66583 Nicandra and 157123 2004NW5.
These young families are now listed in Table 1 and included in the present distribution. The Martes and
Hobson families were mentioned in NRVB24 but were not included in their tables or the total count. With this correction we
have 291 previously known families. With the 63 new cases identified here (see below), 2 new Trojan
families reported in Vokrouhlick\'y et al. (2024b) and 4 new Hilda families from Vokrouhlick\'y et al. (2025),
there are now 360 known asteroid families.\footnote{15156 2000FK38 is not counted here. In addition,
  several very old main-belt families were reported in Delbo et al. (2017, 2019), most notably (161) Athor
  and (689) Zita.}

\subsection{New young families}

Tables 2 and 3 report the new young asteroid families identified in this work. Given our systematic approach to
the problem, this list should be (practically) complete (in the sense that it may be difficult to extract
more real families from the exiting data; but see Section 3.5),\footnote{We also searched for new
  young families in 5D in the osculating element catalog and found a good agreement with families identified
  in 5D in the proper element catalog.} at least for the NRVB24 catalog --
there is no doubt that many more asteroid families will be discovered with future data from the Vera Rubin
observatory. Most newly identified families are small, having only 3-10 members, but there are several exceptions.
The largest family identified in this work is 114555 2003BN44 with 58 members. This family is located in the
much larger Dora family (Fig. \ref{big}; just like the Karin family is located in the much larger Koronis family).
Two nearby asteroids 16472 and 30693 are offset in proper elements and do not seem to be members; no small
family was reported near these bodies in the previous publications. The family is estimated to have formed
$3\pm1$ Myr ago (Fig. \ref{big2}).

The second largest family, totaling 33 members, was found near the inner main-belt asteroid 403307 2009CR6 with $H=18$ mag.
This object is offset from the rest of the family (Fig. \ref{big3}) and does not participate in the orbital
convergence of all other members (403307 is not shown in Fig. \ref{big4}); it is probably an interloper. Finding
this relatively large family was a surprise to us given that the other families listed in Tables 2 and 3, with faint
largest members, typically have only 3 members. Indeed, the brightest family members have $H\simeq18.6$
mag, which does not leave much space -- in terms of the magnitude range -- for many additional members (the faintest
member has $H=20.0$ mag). The new family is a special case, however, with many identified members having a similar
brightness. The family has a very steep size distribution and was probably created by a super-catastrophic
breakup of a relatively large parent body (Michel et al. 2015). This also presents a difficulty in naming the
family. We opted for naming the family after 2021PE78 with $H=18.9$ mag which is located near the family center
(Fig. \ref{big3}). The 2021PE78 family is $1.1\pm0.5$ Myr old (Fig. \ref{big4}).

In addition, there are six new young families with more that 10 members:
5950 Leukippos (16 members, $t_{\rm age}=1.3\pm0.5$ Myr),
5971 Tickell (13 members, $t_{\rm age}=1.5\pm0.3$ Myr),
28805 Fohring (12 members, $t_{\rm age}=0.7\pm0.3$ Myr),
34216 2000QK75 (22 members, $t_{\rm age}=1.5\pm0.7$ Myr),
41331 1999XB232 (18 members, $t_{\rm age}=1.7\pm0.5$ Myr),
and
503256 2015KL76 (14 members, $t_{\rm age}<3$ Myr).
Figure \ref{bigfam} shows the convergence of proper angles for these families.\footnote{The 2961 Katsuharama family
  was identified in NRVB24. The large members of this family show the convergence of proper longitudes at $\simeq 1$ Myr
  ago, suggesting this family is relatively young. We do not include the Katsuharama family in this work because this
  case will need a more detailed study to understand the orbital behavior of small members. This family is also not 
  included in the total count of families.}

As for the very young new families with $t_{\rm age} \leq 1$ Myr, there are 28 families with 3 members,
10 families with 4 members, 7 families with 5 members and 4 families with 6 members, together representing 92\%
of the total number of very young families with $t_{\rm age} \leq 1$~Myr.
It is expected that most new young families should have very few members, because
these member asteroids are faint and at the limit of our current telescopic capabilities. In this sense, the identified
members represent the tip of the iceberg, and many more members will probably be found in the future.
In at least some cases, the new families may have been produced by rotational fission when the parent body split and some of the
fragments became unbound (Pravec et al. 2010). If that is the case, we do not expect (many) additional members to be
discovered in these families (Section 3.6).

\subsection{Family age estimates}

In Section 2.4, we described three different approaches to the family age estimation: the (a) convergence
of proper angles computed from NRVB24, (b) convergence of osculating angles from simulations that ignore
orbital uncertainties and the Yarkovsky effect, and (c) backward integrations of asteroid clones that account for
orbital uncertainties and the Yarkovsky effect. Here we compare these methods for several new young
families (Fig. \ref{prop1}).

The 5722 Johnscherrer family has 4 members and shows the orbital convergence at $\sim 1$~Myr ago. The convergence is
clear for all methods described above.\footnote{The four families discussed here, 5722 Johnscherrer, 7629 Foros,
  8306 Shoko, 10484 Hecht, are located in the inner main belt. For 5722 and 8306, both in the Flora family, we adopted
  the mean albedo of the Flora family in method (c), $p_{\rm V}=0.29$ according to Dykhuis et al. (2014). 
  Asteroids 7629 and 10484 were assigned albedos $p_{\rm V}=0.25$ and $p_{\rm V}=0.23$ (Mainzer et al. 2019),
  respectively. We addopted the physical properties of S-type asteroids all members of the four families
  (Section 2.4).}  
There is not much difference between methods (b) and (c) (i.e.,
with and without the orbital and Yarkovsky clones; the left panels in Fig. \ref{prop1}). This is a consequence of all four members  
of the family having small orbital uncertainties (e.g., the semimajor uncertainty $< 2.1 \times 10^{-8}$
au) and being relatively large ($H=14.3$, 17.2, 18.1 and 18.7 mag) for the Yarkovsky drift to be important. Still,
when longer timescales are considered, the Yarkovsky drift can produce a convergence for $t>1$~Myr.
In this sense, the age determined here would strictly be the {\it lower bound} on the age. We tested this by generating mock
families similar to 5722 Johnscherrer and found that the methods (a) and (b) are able to recover the {\it true} age
quite accurately, and that the minimum age from method (c) often coincides with the real age. Based on these results
we conclude that the age of the 5722 Johnscherrer family is $t_{\rm age}=0.9 \pm 0.3$~Myr. The error bars given here are
conservative (see Fig. \ref{prop1}).

The 7629 Foros family has 3 members and shows the orbital convergence at about 300~kyr ago. There is a relatively good agreement
between methods (a), (b) and (c). This family is located in the inner belt, $a=2.36$ au, and has a substantial
orbital eccentricity, $e_{\rm p}=0.2$. The orbital dynamics in this region is often chaotic due to overlapping Martian
resonances. This may explain
the slight difference between method (a) on one side, and methods (b) and (c) on the other side, because method (a)
is based on proper angles inferred from a 5~Myr integration -- chaotic effects are already noticeable in this longer
interval. We estimate from methods (b) and (c) that the Foros family is $0.3\pm0.1$ Myr old. Method (c) with the
Yarkovsky clones of small family member 2022 SO223 ($H=19.8$ mag) would allow for older ages as well but given our
tests mentioned above it is likely that the true age is near the minimum age from the method (c)
(about 0.3 Myr). Interestingly, 2022 SO223 would need a strong positive Yarkovsky drift in the semimajor axis for
the convergence to happen near 0.3 Myr. This may indicate that 2022 SO223 has a spin state with
the obliquity near 0.

The 8306 Shoko family with 4 members is somewhat similar to 7629 Foros as for the orbital characteristics 
(inner belt, high eccentricity). The largest member of this family is a binary (Pravec et al. 2019a and the
references therein); this family is a good candidate for rotational fission.
Again, there is a slight difference between the convergence of proper angles and
methods (b) and (c). Without accounting for the orbital uncertainty and Yarkovsky drift, the small family
member 2023 SP34 ($H=20.3$~mag) appears to diverge from other family members in nodal longitude but this is
corrected when these effects are accounted for; 2023 SP34 is very likely a member of the family (see Section 3.4).
We investigated this in detail and found that the main cause of differences in the nodal behavior of
2023 SP34 is the relatively large orbital uncertainty. For example, the current 1-sigma uncertainty in the semimajor
axis of 2023 SP34 is $6 \times 10^{-4}$~au, which is significant because it is comparable to the semimajor axis
width of the whole Shoko family ($8 \times 10^{-4}$~au). Our best age estimate for the Shoko family is $t_{\rm age}=0.4\pm0.1$ Myr.

Finally, we examined the 10484 Hecht family (Pravec et al. 2019a;
the right panels in Fig. \ref{prop1}). This family has three relatively large members
($H=14.0$, 15.1 and 18.3 mag) and more stable orbits in the inner belt ($a=2.32$ au, $e_{\rm p}=0.1$).
Two asteroids with nearby orbits, 75630 2000AR51 and 2008UF101, have offset proper longitudes and are probably not
members of the Hecht family. There is a good agreement
between the three methods. The osculating perihelion longitude difference shows large oscillations, which is probably
tied to the relatively low orbital eccentricity of this family. Here it is better to base the family age
estimate to the behavior of nodal longitudes. The minimum age from method (c) is about 0.25 Myr. Our best age estimate
for the 10484 Hecht family is $t_{\rm age}=0.25\pm0.05$ Myr.  

To summarize, there is a generally good agreement between methods (a), (b) and (c). Method (a), which is the most
straightforward to apply, is the least accurate. This method may give inaccurate results for families
that have high orbital eccentricities, because the orbits of their members may not be stable enough to
accurately define the proper angles. Method (c) gives the minimum age and often allows for, especially for very small
family members, the convergence for older ages as well. According to our tests, however, the true age of a family
often falls very close to the minimum age
derived from method (c). Method (b) is a good compromise between complexity and accuracy. Figures \ref{davok} and
\ref{davok2} show the results of method (b) for a dozen new young families. Tables 2 and 3 report our best age estimates
for all new families. 

\subsection{Statistical significance} 

There are 63 new young families in total (Tables 2 and 3). Here we address the statistical significance of these
families. Let us first consider two super compact families reported here (Table 4). From the dispersion of these
families in proper elements, we conservatively estimate that they occupy (fractional) 5D volume that represents 
$\sim 5 \times 10^{-23}$ for 23637 or $\sim 3 \times 10^{-18}$ for 111298 of the total 5D volume available to orbits
of main belt asteroids. Thus, with 1.25 million orbits in total, the probability that the second and third members of
these families fall, by chance, in the same volume element as the first member is
$(1.25\times10^6 \times  5 \times 10^{-23})^2 \sim 4 \times 10^{-33}$ for 23637 and $\sim 10^{-23}$ for 111298. The probability
that this happens once by chance for any of 1.25 million orbits is
$1.25\times10^6 \times 4 \times 10^{-33} \sim 5 \times 10^{-27}$ for 23637 and $\sim 10^{-17}$ for 111298. These two families
are obviously statistically significant.\footnote{Many of the new young families reported
  in this work are located in the previously
known background families (see Notes in Tables 2 and 3), where the orbital
density of background asteroids is larger than the main belt average. For
example, 23637 is located in the densely populated Vesta family. This raises
a question of how the statistical significance of new families could be affected
by this. As an example, we estimate that 23637 represents the fractional 5D
volume $\sim 5 \times 10^{-20}$ of the Vesta family, which currently has
$\sim 3\times10^4$ members. So, repeating the calculation from the main text,
we find that 23637 has the $\sim 10^{-25}$ chance to occur once in the Vesta
family. This can be compared to $\sim 5 \times 10^{-27}$ for 23637 to happen
once in the whole main belt.} 
Figure \ref{davok3} shows our convergence tests for some of the most compact families found in this work. 

We repeated the same estimate for all families reported here and found that all families with more then three members
are statistically significant (adding additional members enormously increases the statistical
significance; NRVB24). Some of the least compact families with three members reported in Tables 2 and 3,
such as 153093 or 208804, occupy the fractional volume $<10^{-12}$. The above logic applied to these families gives a
$< 2 \times 10^{-6}$ probability that this could happen once, by chance, in the whole asteroid belt.
These families are thus clearly significant as well. Other families reported in Tables 2 and 3 are intermediate between
23637/111298 and 153093/208804. We thus conclude that all cases reported in Tables~2 and 3 are {\it real} families.  

The orbital distribution of young families in the asteroid belt is shown in Fig. \ref{orb}. Most of the new young
families identified here are located in the inner belt. This makes sense because the new families often have very
faint members which are detectable by telescopic observations only when they are relatively close to the
terrestrial observer. Five new families were found in the Hungaria region. The middle and outer asteroid belt
are thought to represent $\sim 2.5$-3 times and $\sim 5$-6 times larger populations than the inner asteroid belt
(Masiero et al. 2011). This means, even if the telescopic surveys bias the population statistics toward the inner
belt, that there should in reality be more young families in the middle and outer belts. Kurlander et al. (2025)
estimated that the Vera Rubin observatory should detect $\simeq 99$\% of main belt asteroids with red magnitude
$H_{\rm r}<18.5$ and $\simeq80$\% with $H_{\rm r}<19$. We therefore expect the number of young, middle/outer asteroid
belt families to significantly increase. 

Interestingly, many of the new young families are members of known older families. These associations are listed
in Tables 2 and 3. Overall, about 54\% of the new families are located in old families and 46\% is located in the
background. For some reason, only two of the new young families are members of older C-type families, 114555 in
Dora and 237295 in Clarissa; all others are in the S-complex families.

\subsection{Completeness}

These estimates raise questions related to the completeness of the young family catalog provided here. If all families
reported here have a very high statistical significance, would it be possible to relax our family identification
criteria and identify new families that are not as highly significant as the ones in Tables 2 and 3, but still significant enough?
For example, we imagine a situation where two or more large fragments are ejected from a parent body with large ejection speeds;
this would lead to large differences $\delta$ between proper orbits in Eq.~(1), and would require the use of a
larger cutoff $d_{\rm cut}$.\footnote{A good example of this is the 63440 2001MD30 family, previously identified in Pravec
  et al. (2019a) and Fatka et al. (2020), that our detection algorithm from Section 2.3 missed. This young family is
  slightly more dispersed than other families identified here and requires $d_{\rm cut}=20$ m s$^{-1}$ (instead of
  the standard 10 m s$^{-1}$).}     
With the larger cutoff, however, the rate of false positives would increase and it would
be more difficult to distinguish between what is real and what is not. We leave this investigation for future
work and only note here that this effort could be important for the identification of massive cratering events
on relatively large parent bodies, for which the ejection speeds could be relatively high.

In addition, with our strict identification criteria, we may have missed families with $t_{\rm age}\gtrsim1$ Myr, especially
the ones with small members that may have accumulated substantial Yarkovsky drift over the family age. By ignoring this
drift in the identification method (Section 2.3), we may have not detected a tight convergence of $\Omega_{\rm p}$ and $\varpi_{\rm p}$,
if the family is older than some threshold.\footnote{A good example of this is the 157123 2004NW5 family, previously identified
  in Pravec et al. (2019a) and Fatka et al. (2020), that our detection algorithm from Section 2.3 missed. When integrated
  backward without the Yarkovsky effect, the small members of the family do not show any obvious convergence of angles. 
  The family can be identified by the standard HCM in 3D with $d_{\rm cut}=20$ m s$^{-1}$.}
To address this issue, we plot in Fig. \ref{ages} the ages of young families
listed in Tables 1--3. There are several notable trends in this plot. First of all, the identified young families
with $t_{\rm age}>2$ Myr typically have many members and a relatively bright largest member. The families with
many members are apparently easier to identify, even if they formed more than 2 Myr ago.
It is also easier in these cases to reliably estimate the family age.

Second, none of the very young families with $t_{\rm age}<2$ Myr have
a bright largest member with $H<12$ mag  (rectangles A and B in Fig.~\ref{ages}), except for our new 1346 Gotha family.
This is probably a real feature, unrelated to detection biases, because fewer very large cratering events or
catastrophic breakups
are expected to happen in the last 2 Myr. It therefore
makes sense that the parent bodies of identified very young families are relatively small. 
With $d_{\rm cut} \sim 10$~m~s$^{-1}$ (Section 2.3), 
we may have also failed to identify some cratering impacts on large parent bodies.
Third, there are fewer identified families with $t_{\rm age}>1$ Myr (rectangles B and C
in Fig. \ref{ages}) than $t_{\rm age}<1$ Myr (rectangle A).
This is most likely related to a bias of the detection method, because we do not expect to recover the convergence
for small family members that accumulated a substantial Yarkovsky drift, if $t_{\rm age}>1$ Myr. The set of young families
reported here with $t_{\rm age}>1$ Myr is therefore largely incomplete. We did not find any new young families with
$t_{\rm age}>3$ Myr. 

Another obvious bias is related to the absolute magnitude of the brightest family members (Fig. \ref{deltah}).
We find, with only five exceptions shown in Fig. \ref{deltah}, that the second brightest member in a family is always
brighter than $H=19.0$ mag (the sensitivity limit of the ongoing asteroid surveys). This bounds
the magnitude difference between the brightest and second brightest family members, $\Delta H = H_{\rm 2nd}-H_{\rm 1st}$
as a function of $H_{\rm 1st}$, with brighter first family members allowing for larger $\Delta H$.

We recall that the identification of asteroid families in this work is based on the February 2024 MPC catalog, which
was the source of the proper elements computed in NRVB24. As new faint asteroids are added to the MPC catalog, new
families will likely to be identified. As an example, consider the 25435 1999WX3 family. Here we found that
25435 1999WX3 ($H=15.17$ mag) and 2009 SD429 ($H=19.52$ mag) form a very compact asteroid pair with tightly
clustered osculating elements (Table~5). When we check on the most recent release of the MPC catalog, we identify
a potential third member, 2019 SY257, with $H=21.34$ mag. The nominal orbit of 2019 SY257 appears to fall
extremely close to that of 25435 1999WX3 and 2009 SD429, indicating that 2019 SY257 can indeed be a new family member
(Table 5). The current orbital uncertainties of 2019 SY257 are relatively large, however, suggesting caution.
By looking into this issue in more detail we identified another $\sim 5$ cases, where a third member was added in
the 2025 MPC catalog to a pair identified from the 2024 MPC catalog. These cases are not reported in Tables 2 and 3.  

\subsection{Collisions vs. rotational fission}

There is no doubt that a great majority of {\it large} asteroid families were produced by impact cratering and disruptive
collisions. These families often have very large parent bodies, which are not susceptible to spin-up by YORP, and
hundreds to tens of thousands of members, which would require an implausibly large number of fission events.
In contrast to that, rotational fission is thought to be the main source of asteroid pairs (Pravec et al. 2010).
Here we consider the question of the relative importance of collisions and rotational fission for small/young
families identified here (many of which currently have only 3 members; Tables 2 and 3).

The number of family members is a possible metric to distinguish between the two formation mechanisms. Here, a large
impact on a parent body can produce a family with many members. Compared to that, the rotational fission is
relatively inefficient and is expected to produce families with a small number of members. Historically, there were four small
asteroid families identified back in 2006 (Nesvorn\'y \& Vokrouhlick\'y 2006, Nesvorn\'y et al. 2006). Of these, the Datura and
Emilkowalski families now have 77 and 17 members in Table 1, respectively, whereas the Brugmansia (formerly 1992YC2) and
Lucascavin families remained with 3 member asteroids each.\footnote{Over a hundred Datura family members can be
  identified in the newest MPC catalog (June 2025). The Brugmansia family may now have four members.}
The Brugmansia and Lucascavin families would thus be good candidates for rotational fission (see Pravec et al. 2018 for
other examples). Another good candidate is the Kap’bos family, for which Fatka et al. (2020) found two convergence
events significantly separated in time. Given that the upcoming LSST program
at the V. Rubin observatory should discover $\sim 5$ million main belt asteroids (Kurlander et al. 2025), it will be
interesting to see how this surge will affect the membership of young families (Tables 1-3).

The new candidates for rotational fission include the following triples:
14155--437384--631600, 30301--205231--2007RV377, 80245--540161--2017AB57, 100416--2013SJ104--2014PX11, 
100440--575395--2016QS128, 111298--457548--2017VL47, 133303--458905--2021QX55, and 141906--398383--2007TN469
(Tables 2 and 3). Related to that, several of the new young families were previously identified as asteroid pairs (Pravec
et al. 2019a), including 8306 Shoko, 16126 1999XQ86 (6 members now), 30301 Kuditipudi, 46162 2001FM78, 51866 2001PH3,
and 100440 1996PJ6. 

The $P_1$-$\Delta H$ correlation could be another useful tool (Pravec et al. 2010, 2018). If a very large secondary
forms by rotation fission, with the mass ratio $M_2/M_1\gtrsim0.3$, where $M_1$ and $M_2$ are the primary and secondary
masses, the secondary cannot escape because there is not enough free energy in the system to allow for that (even
if all primary's rotational energy is transported to secondary's orbit). The escape is possible for $M_2/M_1\lesssim0.3$.
In that case, for a single fission event, one expects a correlation between $P_1$ and $\Delta H$, with the spin
rate increasing with $\Delta H$ (small secondaries can escape more easily, primary's rotation does not need to slow
down much; Pravec et al. 2010). For more fission events, the above argument presumably applies only to the last one:
the primary's spin state must have evolved by YORP after the previous fission events to allow for the new ones.  
This complicates inferences about the origin of young families. 

Additional arguments can be based on the existence/absence of satellites around the largest family member, largest
member's shape (rounded or not), and the number of identified events with different (estimated) ages (Fatka et al. 2020).

Collisions and rotational fission scale differently with asteroid size. If the cumulative size distribution of main
belt asteroids is approximated by $N(D) \propto D^{-\alpha}$, with some power index $\alpha \simeq 2$-2.5 (Nesvorn\'y
et al. 2024b), the number of fission events is expected to increase as $D^{-\beta}$ with $\beta=\alpha+2$ (here we
assume that the YORP timescale is $\propto D^2$, Vokrouhlick\'y et al. 2015). For comparison, the number of
disruptive collision events scales with $N(D) N(d) D^2$, where $D$ is the target diameter, $d$ is the projectile
diameter, and $D^2$ stands for collisional cross-section. For $d \propto D$ (Benz \& Asphaug 1999; $Q^*_{\rm D}$ scaling
with $D$ represents only a small correction in the strength regime), we find that the number of catastrophic collisions
is expected to increase as $D^{-\gamma}$ with $\gamma=2 \alpha-2$. For $\alpha=2$-2.5, we have $\beta=4$-4.5 and
$\gamma=2$-3. We therefore see that rotational fission events have steeper scaling with asteroid size and should
therefore become more dominant for small asteroid sizes. Marzari et al. (2011) suggested that rotational fission should
become dominant for parent bodies with $D\lesssim2$ km (their Fig. 9).  

\subsection{Constraints on the collisional evolution}

Given the various detection biases discussed above, we consider the question of how the (biased) dataset of young asteroid
families could be used to constrain the collisional evolution of the asteroid belt (Bottke et al. 2005, 2020).
We make a tentative assumption in this section that the majority of young asteroid families in Tables 1-3
were produced by collisions. See the previous section for a discussion of the relative importance of collisions
and rotational fission.

Ideally, a rigorous approach to this problem would require: (1) running sets of collisional models, (2) accounting for
the detection bias, and (3) comparing the biased sets of model families with the catalog provided here.
For example, as for (2), we could account for the
asteroid detection probability by the Catalina Sky Survey (Christensen et al. 2012) and apply HCM to the biased
collisional model to recover a biased set of model families. This approach would represent a substantial work effort.
Alternatively, we could use the collection of larger families shown in Fig. \ref{ages}, because we know that this
sub-sample is essentially unbiased (families similar to the Karin family can easily be identified).
The third possibility would be to use the subset of small, very young families. For example, there are 46 
families with $t_{\rm age}<1$ Myr and $12<H<17$ mag (rectangle A in Fig. \ref{ages}).
A calibrated collisional model should thus produce $\sim 46$ families with these characteristics.
The proposed collisional modeling would be useful to estimate the size distribution of very small
main belt asteroids -- projectiles that produced the small/young families -- and provide
constraints on the impact-scaling laws (e.g., Benz \& Asphaug 1999). We leave this project for future work.

For reference, we estimated the frequency of cratering and catastrophic breakups in the asteroid belt. The specific
energy of a catastrophic breakup, $Q_{\rm D}^*$, when half of the target is dispersed in space, was taken from Bottke
et al. (2020), who inferred $Q_{\rm D}^*$ from modeling the size distribution of main belt asteroids. First, for a
target asteroid of radius $R_{\rm tar}$, we computed the impactor radius, $R_{\rm imp}$, such that $Q/Q_{\rm D}^*=1$,
where $Q$ is the specific energy of impact (the impact speed was set to 5.3 km s$^{-1}$; Bottke et al. 1994). Second,
from the size distribution of main belt asteroids given in Bottke et al. (2020), we estimated the number of bodies
with $R>R_{\rm tar}$ and $R>R_{\rm imp}$, obtaining $N_{\rm tar}$ and $N_{\rm imp}$, respectively. Third, the characteristic
timescale of catastrophic impacts, $\tau$, was obtained from
\begin{equation}
1/\tau = P_{\rm i}\; N_{\rm tar} N_{\rm imp} (R_{\rm tar}+R_{\rm imp})^2 
\end{equation}  
with the intrinsic impact probability $P_{\rm i}=2.9 \times 10^{-18}$ km$^{-2}$ yr$^{-1}$ (Bottke et al. 1994).
The size of the largest fragment
produced by a catastrophic breakup was estimated from Eq. (10) in the Supplementary Materials in Morbidelli et
al. (2009). Finally, for a reference albedo of $p_{\rm V}=0.15$, we computed the absolute magnitude of the largest
fragment, $H_{1st}$, and plotted $\tau(H_{1st})$ in Fig.~\ref{ages}. A similar estimate was done for the cratering
collisions with $Q/Q_{\rm D}^*=0.1$.

In Fig. \ref{ages}, all families to the left of the $Q/Q_{\rm D}^*=1$ line have younger formation ages that the
characteristic time for one catastrophic breakups in the main belt. This means, as a plausibility check, that most of
these families should have been produced by {\it sub-catastrophic} breakups. This is indeed the case. For example, there are
six families below the $Q/Q_{\rm D}^*=1$ line with $H_{\rm 1st}<13.0$ and $t_{\rm age}<1$ Myr: 525 with $\Delta H = 6.2$,
1270 with $\Delta H = 4.2$, 2384 with $\Delta H = 2.8$, 4765 with $\Delta H = 5.5$, and two new families, 1346 with
$\Delta H = 6.1$ and 5478 with $\Delta H = 5.0$. These large brightness differences between the first and second
largest fragments imply, with a possible exception of 2384, the sub-catastrophic breakups (Fig. \ref{deltah}).
In addition, the new 1346 Gotha family with $H_{\rm 1st}=11.5$ and $t_{\rm age}=0.6 \pm 0.1$ Myr is also below
the $Q/Q_{\rm D}^*=0.1$ in Fig. \ref{ages}, but this is consistent with this family corresponding to a large cratering
event on 1346 Gotha ($\Delta H = 6.1$; Fig. \ref{deltah}).  

\section{Conclusions}

\begin{figure}
\epsscale{0.6}
%\plotone{bigf2.JPG}
\plotone{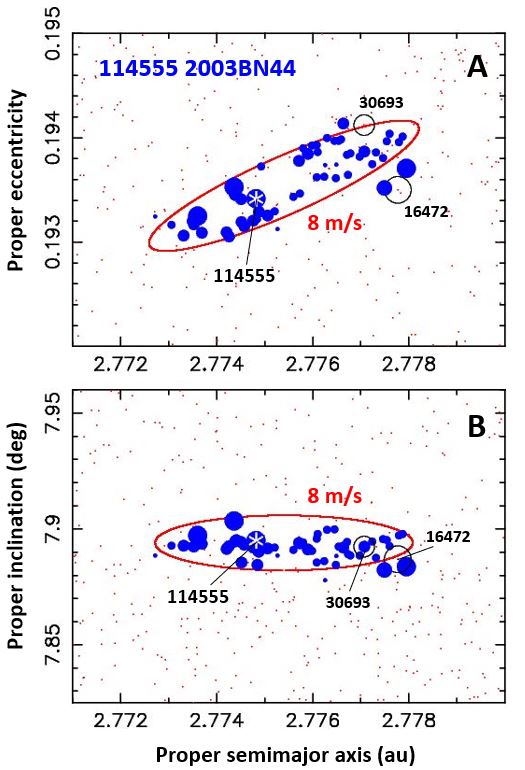}
\caption{114555 2003BN44 is the largest young family identified in this work. It has 58 members identified
  by 3D HCM with $d_{\rm cut}=7$ m s$^{-1}$. Two particularly large HCM members, 16472 and 30693, are offset
  from the family center and may be interlopers. The red ellipses plotted here were computed from the
  Gauss equations with the ejection speed 8 m s$^{-1}$, true anomaly $f=60^\circ$ and perihelion
  argument $\omega=0$ (Nesvorn\'y et al. 2002).}  
\label{big}
\end{figure}

\begin{figure}
\epsscale{0.55}
%\plotone{114555_2003BN44.eps}
%\plotone{big3.JPG}
%\plotone{snip15.JPG}
\plotone{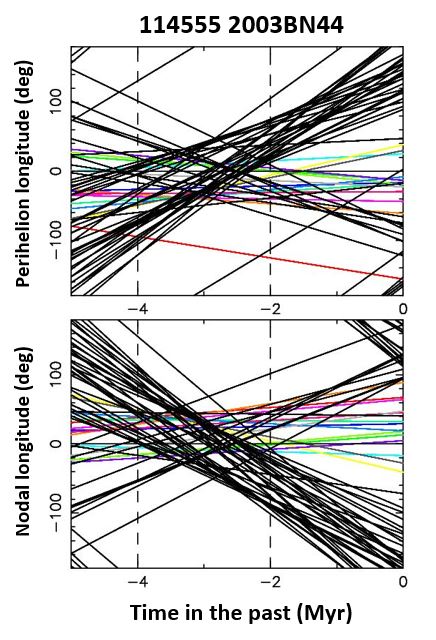}
\caption{The convergence of proper angles indicates that the 114555 2003BN44 family formed $3 \pm 1$ Myr ago.}
\label{big2}
\end{figure}

\begin{figure}
\epsscale{0.6}
%\plotone{snip31.JPG}
\plotone{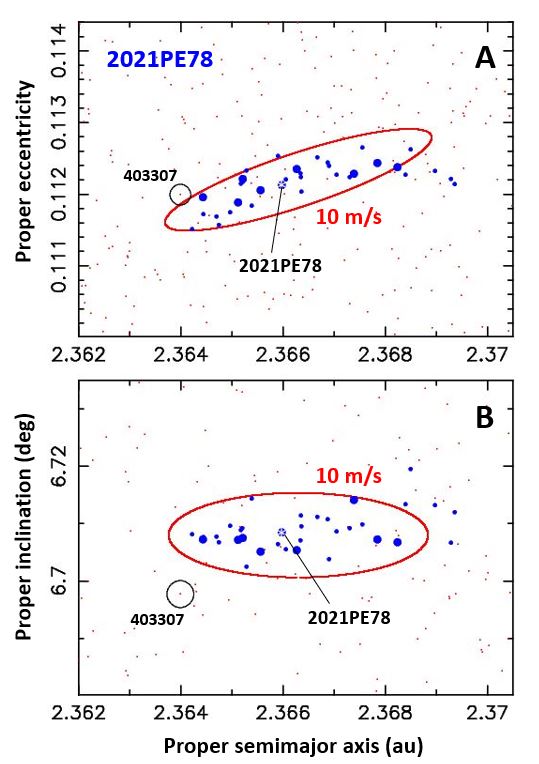}
\caption{2021PE78 is the second largest young family identified in this work. It has 35 members identified
  by 3D HCM with $d_{\rm cut}=10$ m s$^{-1}$. One particularly bright HCM member, 403307 with $H=18$ mag, is
  offset from the rest of the family and is probably an interloper. Another HCM member, 2017QV170, does
  not participate in the orbital convergence and may be an interloper as well. This leaves 33 members. 
  The red ellipses plotted here were computed from the Gauss equations with the ejection speed 10 m s$^{-1}$,
  true anomaly $f=60^\circ$ and perihelion argument $\omega=15^\circ$ (Nesvorn\'y et al. 2002).}  
\label{big3}
\end{figure}

\begin{figure}
\epsscale{0.55}
%\plotone{114555_2003BN44.eps}
%\plotone{big3.JPG}
%\plotone{snip30.JPG}
\plotone{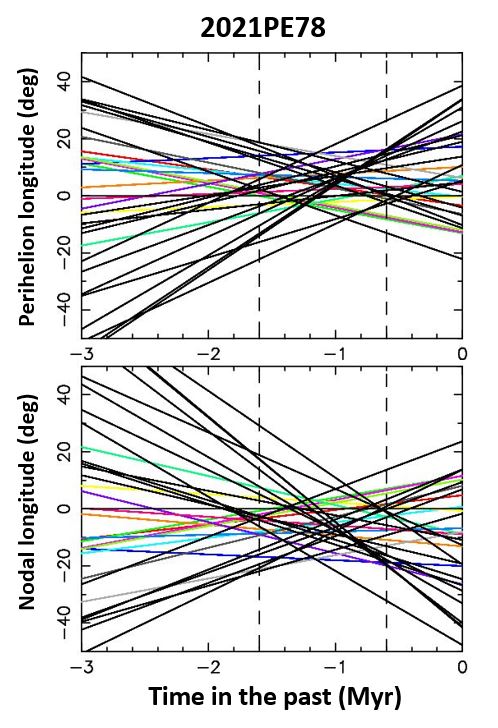}
\caption{The convergence of proper angles indicates that the 2021PE78 family formed $1.1 \pm 0.5$ Myr ago.}
\label{big4}
\end{figure}

\begin{figure}
\epsscale{1.0}
%\plotone{conv_5950.JPG}
%\plotone{conv_5971.JPG}
%\plotone{conv_28805.JPG}
%\plotone{conv_41331.JPG}
%\plotone{big4.JPG}
\plotone{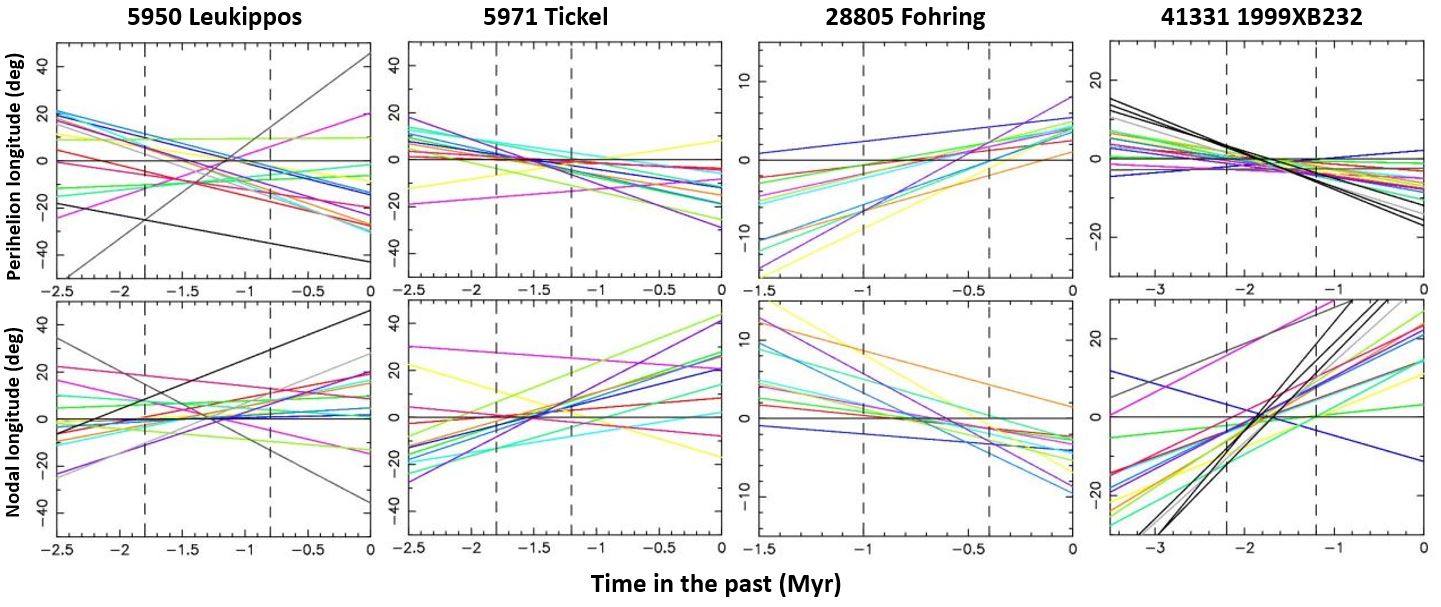}
\caption{The convergence of proper angles for some of the larger young families found in this work: 5950 Leukippos,
  5971 Tickel, 28805 Fohring, 41331 1999XB232. All members identified by 3D HCM are plotted
  ($d_{\rm cut}=10$ m s$^{-1}$). See Table 2 for the number of family members in each case. 
  A few of the nominal members that do not participate in the convergence may be interlopers.
  The vertical dashed lines delimit the admissible age interval.}
\label{bigfam}
\end{figure}

\begin{figure}
\epsscale{0.8}
%\plotone{bigobr5.JPG}
%\plotone{bigobr4.JPG}
\plotone{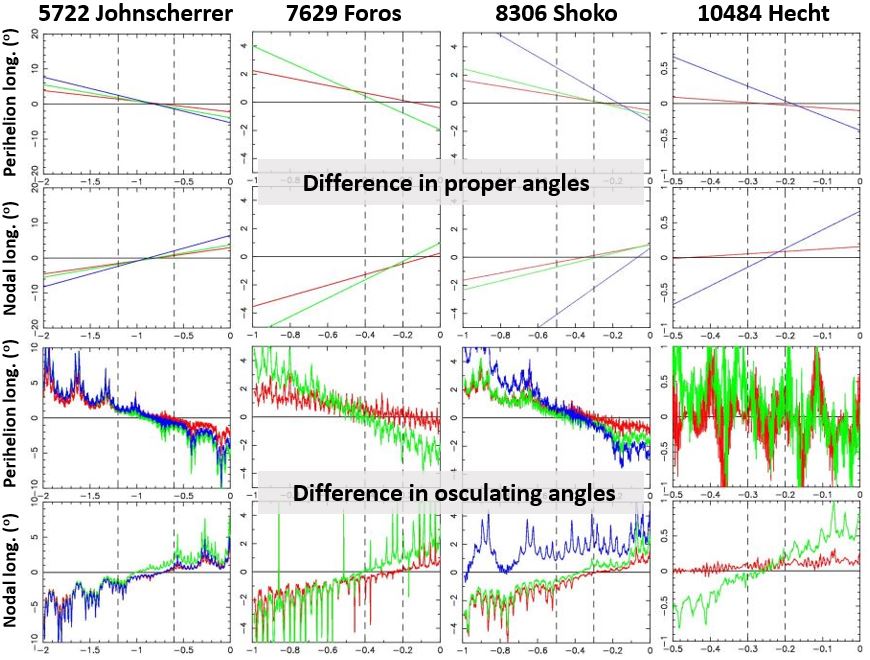}
\plotone{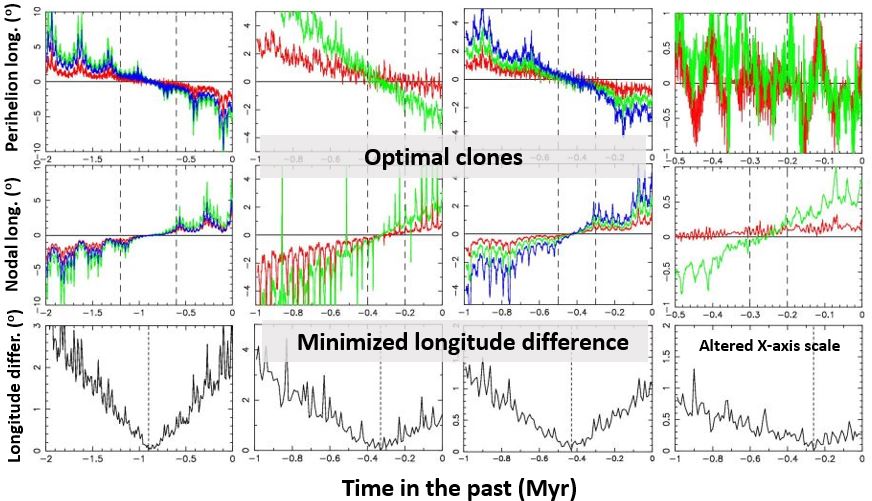}
\caption{\footnotesize A comparison of different age-determination methods (Section 2.4)
  for four young families.
  The top row shows differences in the {\it proper} perihelion longitudes and {\it proper} nodal longitudes
  (method a). The second row shows differences in the osculating longitudes computed from backward integrations that
  ignored the orbital uncertainty and Yarkovsky drift of orbits (method b). The third row shows differences in
  the osculating longitudes from integrations with 110 clones of each family member (method c).
  The orbital evolution shown here corresponds to clones with the optimized convergence.
  The combined (average) difference in longitudes for these clones is shown in the last row. } 
\label{prop1}
\end{figure}

\begin{figure}
\epsscale{0.8}
%\plotone{snip3.JPG}
\plotone{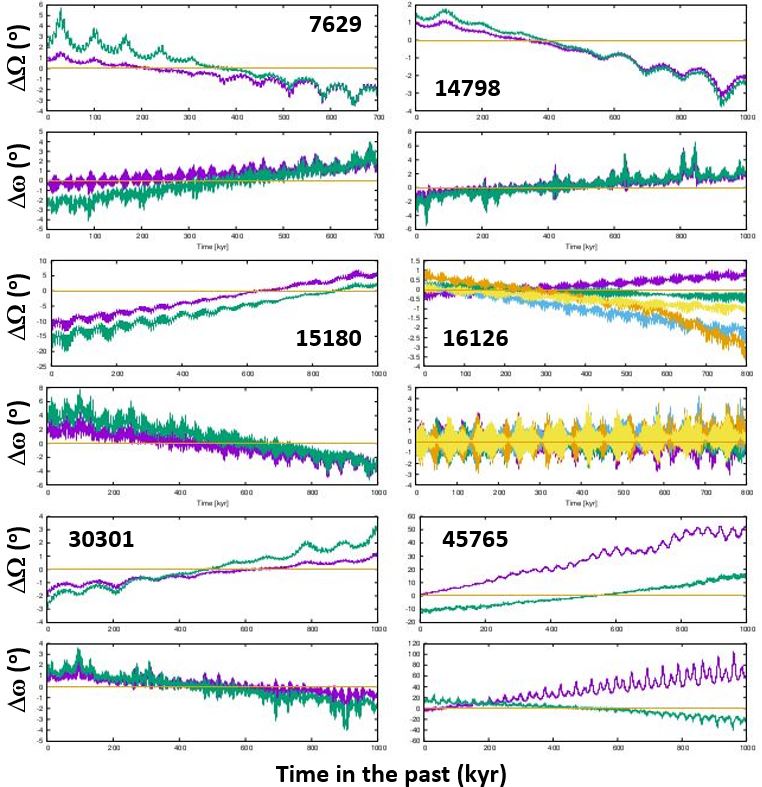}
\caption{The orbital convergence tests for six new young families: 7629, 14798, 15180, 16126, 30301,
  45765. The differences in the osculating nodal
  longitudes ($\Delta \Omega$) and osculating perihelion longitudes ($\Delta \varpi$) are shown
  here. Unlike in the previous figures the clock rewinds back in time from the left to the right. }
\label{davok}
\end{figure}

\begin{figure}
\epsscale{0.8}
%\plotone{snip4.JPG}
\plotone{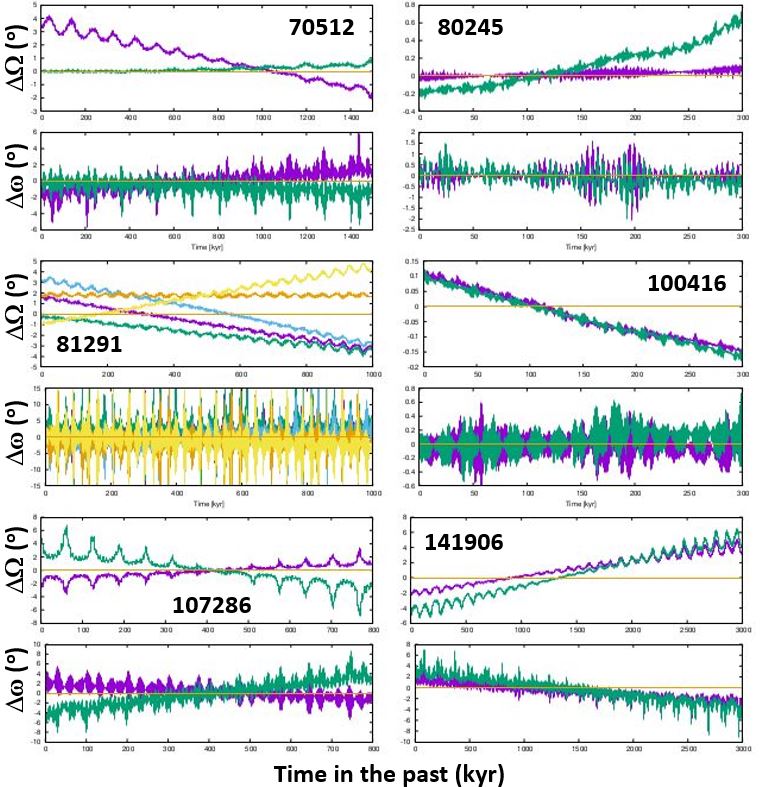}
\caption{The orbital convergence tests for six new young families:
  70512, 80245, 81291, 100416, 107286, 141906. The differences in the osculating nodal
  longitudes ($\Delta \Omega$) and osculating perihelion longitudes ($\Delta \varpi$) are shown
  here. The clock rewinds back in time from the left to the right.}
\label{davok2}
\end{figure}

\begin{figure}
\epsscale{0.8}
%\plotone{Figs/25435.eps}
%\plotone{Figs/100440.eps}
%\plotone{Figs/180023.eps}
%\plotone{Figs/267721.eps}
%\plotone{davok3.JPG}
\plotone{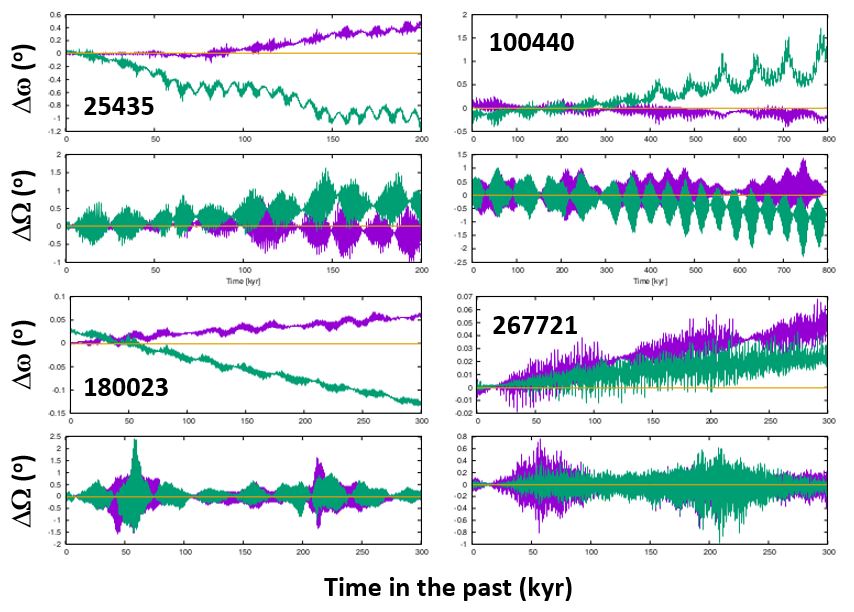}
\caption{The orbital convergence tests for four very compact families: 25435, 100440, 180023 and
  267721. The differences in the osculating nodal longitudes ($\Delta \Omega$) and osculating perihelion
  longitudes ($\Delta \varpi$) are shown here. The clock rewinds back in time from the left to the right.
  The 25435 family was identified as a pair in the NRVB24 catalog. The third member of this family
  was found in the recent release of the MPC catalog.}
\label{davok3}
\end{figure}

\begin{figure}
\epsscale{0.8}
%\plotone{update7.JPG}
\plotone{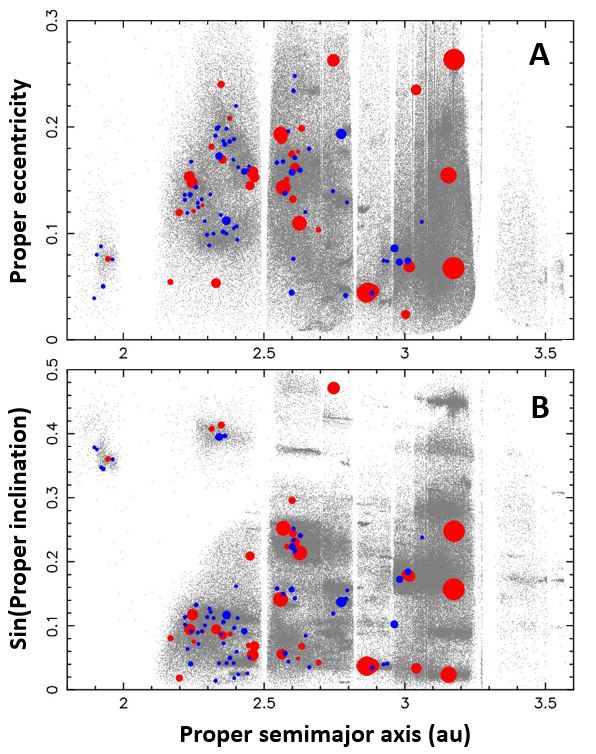}
\caption{The orbital distribution of young asteroid families. The symbol size is proportional
  to the number of family members. The color indicates whether the family was known previously (red)
  or is new (blue). The main belt asteroids from the NRVB24 catalog are plotted in the background.
  We only plot background asteroids with $H<16.3$ -- this population is complete according to Hendler \& Malhotra
  (2020).}
\label{orb}
\end{figure}

\begin{figure}
\epsscale{0.85}
%\plotone{update2.JPG}
\plotone{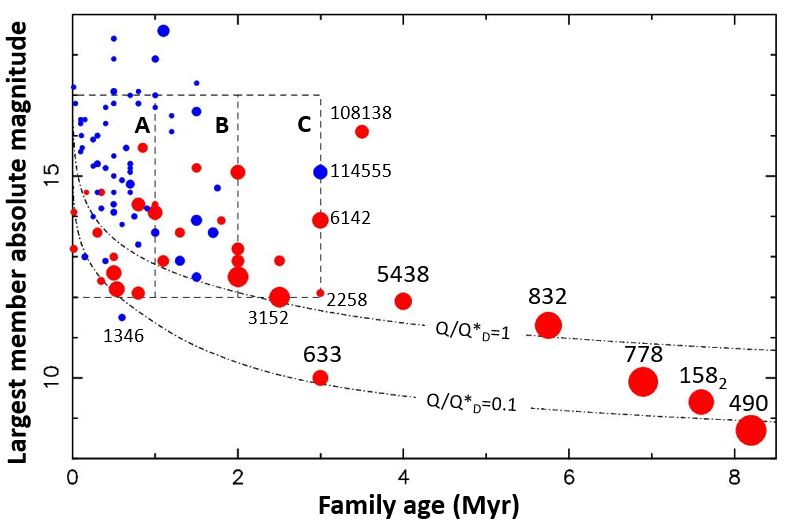}
\caption{The age estimates for young asteroid families and absolute magnitudes of their brightest members
  (Tables 1--3). The symbol size is proportional to the number of family members. The color
  indicates whether the family was known previously (red) or is new (blue). For families in
  Tables 1--3 where only an upper age bound is available, half of the upper bound is plotted here.
  Rectangles B and C contain fewer families than rectangle A, suggesting that
  the known sample of families with faint largest members becomes increasingly incomplete
  with age. The dash-dotted lines show our estimates for the frequency of catastrophic
  ($Q/Q^*_{\rm D}=1$) and cratering ($Q/Q^*_{\rm D}=0.1$) collisions in the asteroid belt.}
\label{ages}
\end{figure}

\begin{figure}
\epsscale{0.9}
%\plotone{update4.JPG}
\plotone{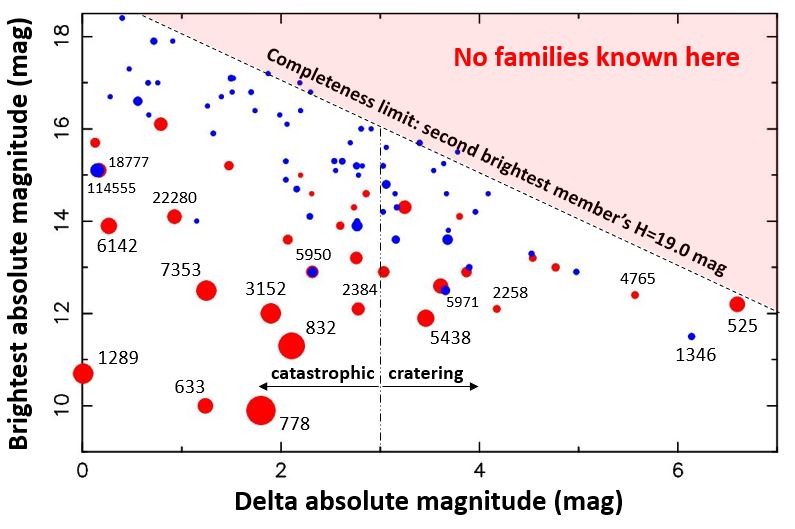}
\caption{The absolute magnitude of the brightest family member, $H_{\rm 1st}$, as a function of  
  the absolute magnitude difference between the brightest and second brightest family members,
  $\Delta H = H_{\rm 2nd}-H_{\rm 1st}$. The symbol size is proportional to the number of family members.
  The color indicates whether the family was known previously (red) or is new (blue). All known
  young asteroid families, except for five new families in the inner belt and Hungarias,
  have $H_{\rm 2nd}<19.0$ mag. For a simple fragmentation model, Eqs. (10) and (11) in the
  Supplementary Information of Morbidelli et al. (2009), the catastrophic ($Q/Q_{\rm D}^*>1$)
  and cratering ($Q/Q_{\rm D}^*<1$) impacts would correspond to $\Delta H \lesssim 3$ mag and
  $\Delta H \gtrsim 3$ mag, respectively, as indicated by the vertical dash-dotted line.}
\label{deltah}
\end{figure}

The main results of this work are summarized as follows.
\begin{enumerate}
\item We discovered 63 young asteroid families with the formation ages $t_{\rm age}<10$ Myr. There were 291 previously known
  families (Nesvorn\'y et al. 2015, 2024a; the references therein and Section 3.1). With 63 new cases, 2 new Trojan families reported
  in Vokrouhlick\'y et al. (2024b) and 4 new Hilda families from Vokrouhlick\'y et al. (2025), there are now 360 known asteroid
  families in total. 
\item Three 
  convergence methods were applied to establish the most accurate formation age of each new family (Tables 2 and 3).
  We also revised the age of several previously known families (Table 1). The great majority of young families have a
  relatively small brightest member ($12<H<17$ mag) and $t_{\rm age}<2$ Myr (Fig. \ref{ages}).
\item The largest new families discovered here are 114555 2003BN44 with 58 members (Fig.~\ref{big}) and 2021PE78
  with 33 members (Fig. \ref{big3}). These families are estimated to have formed $3\pm1$ Myr ago (Fig. \ref{big2})
  and $1.1\pm0.5$ Myr ago (Fig. \ref{big4}), respectively.
\item  There are eight new young families with more than 10 members, including 5950 Leukippos (16 members,
  $t_{\rm age}=1.3\pm0.5$ Myr), 5971 Tickell (13 members, $t_{\rm age}=1.5\pm0.3$ Myr), 28805 Fohring (12 members,
  $t_{\rm age}=0.7\pm0.3$ Myr), 34216 2000QK75 (22 members, $t_{\rm age}=1.5\pm0.7$ Myr), 41331 1999XB232 (18 members,
  $t_{\rm age}=1.7\pm0.5$ Myr), and 503256 2015KL76 (14 members, $t_{\rm age}<3$ Myr).
\item In total, there are 46 known families with $t_{\rm age}<1$ Myr and $12<H<17$ mag. These families
  provide important constraints on the collisional evolution of the asteroid belt. A calibrated collisional model should
  produce $\sim 46$ families with these characteristics.
\item  In at least some cases, the new families with a small number of members (3 or 4) may have been produced by rotational
  fission (Pravec et al. 2010). A good example of this is the previously known Lucascavin family (Vokrouhlick\'y et al. 2024a).
  The new candidates for rotational fission are listed in Section 3.2.
\item About 54\% of the new families are located in old families and 46\% is located in the
  background. For some reason, only two of the new young families are members of older C-type families, 114555 in
  Dora and 237295 in Clarissa; all others are in the S-complex families.
\end{enumerate}  

\acknowledgements

\begin{center}
{\bf Acknowledgments} 
\end{center} 
\vspace*{-3.mm}
The simulations were performed on the NASA Pleiades Supercomputer, and D.N.'s and D.V.'s personal workstations.
We thank the NASA NAS computing division for continued support. The work of D.N. was funded by the NASA SSW program.
D.V. and M.B. acknowledge support from the grant 25-16507S of the Czech Science Foundation. F.R. acknowledges support
from the Brazilian Council of Research (CNPq) through grant no. 312429/2023-1.
We thank Petr Pravec and an anonymous reviewer for excellent reviews of the submitted manuscript.

\begin{table}
\label{tab1}  
\centering
{ \scriptsize
\hspace*{-0.8cm}
\begin{tabular}{rllllll}
\hline \hline
Num. & Name & $H_{\rm 1st}$ & $d_{\rm cut}$ & \# of   & $t_{\rm age}$  & Notes \\
     &      & (mag)   & (m/s)  & mem.   & (Myr) &      \\
\hline
158  & Koronis$_2$    & 9.4  & 10    & 1380 & $7.6\pm0.2$       & double $\sin i$-cut, MolHae08, Nes+15, Bro+24a \\ %157
321 & Florentina      & 10.2 & 10    & 209 &  --               & $\sin i$-cut, Koronis$_4$ in Bro+24a, Nes+24\\ % 318
490 & Veritas         & 8.7  & 20    & 6375 & $8.3\pm0.1$      & Nes+03, Tsi+07, Car+17, Bro+24b\\ % 484
525 & Adelaide        & 12.2 & 10    & 86  & $0.536\pm0.012$   & NovRad19, Vok+21a+24a, not listed in Nes+24\\ % 518
%623 & Chimaera        & 10.8 & --    & --  &  --               & not real?, Mas+13, Nes+15 \\ % 614
633 & Zelima          & 10.3 & 10    & 88  & $\sim 3$          & in Eos, z$_1$, unclear convergence, Tsi19, CarRib20 \\ % 623
656  & Beagle         & 10.1 & 20    & 638 & --                & Nes+08, Car19, Nes+24, Beagle interloper\\ %646 not for HCM
778  & Theobalda      & 9.9  & 30    & 4848 & $6.9\pm2.3$      & Nov10\\ % 764
832  & Karin          & 11.3 & 40    & 2201 & $5.75\pm0.05$    & HCM in 5D, Nes+02, NesBot04, Car+16 \\ % 818
1217 & Maximiliana    & 12.9 & 20    & 28  &  0.7-1.5*          & Nes+24\\% 1186  
1270 & Datura         & 12.6 & 10    & 77  &  0.45-0.6         & Nes+06c, Vok+09, Vok+17+24a\\% 1235  
1289 & Kutaisii       & 10.7 & 10    & 371 &  --                 & Koronis$_3$ in Bro+24a\\% 1254
2110 & Moore-Sitterly & 13.6 & 10   & 17  & 1.2-1.5           & PraVok09, Pra+19a, Nes+24\\% 2023
2258 & Viipuri        & 12.1 & 10    & 8    & 2.5-3.5*          & NovRad19, not listed in Nes+24\\% 2161
2384 & Schulhof       & 12.1 & 10    & 45  &  $0.8\pm0.2$      & PraVok09, Vok+11+16+24a\\% 2280
3152 & Jones          & 12.0 & 30    & 341 &  $2.5\pm0.5$*     & Fig. \ref{case1}, Nes+15, Car+18a\\ % 3030
4652 & Iannini        & 13.5 & 20    & 3110 & $6 \pm2 $         & also Nele, Nes+03, Car+18b, Bro+24b\\% 4445
4765 & Wasserburg     & 14.0 & 10    & 8   &  0.2-0.5*          & VokNes08, Pra+19a, Nov+22, Vok+24a, not listed \\% 4549
5026 & Martes         & 14.1 & 10    & 6   &  $0.018\pm0.001$  & VokNes08, Pra+19a, Vok+24a, not listed \\% 4787
5438 & Lorre          & 11.9 & 20    & 150   &  2.0-6.0*      & Nov+12, Nes+15,\\% 5164 
5478 & Wartburg       & 12.9 & 10    & 5    & $0.4\pm0.2$*     & 3 members in Pra+19a, not listed\\ %5202 \\
6084 & Bascom         & 13.0 & 10    & 10  &  0.2-0.8*          & binary, Hig+06, Pra+12a,b, Nes+24\\% 5768
6142 & Tantawi        & 13.9 & 30    & 114 &  2.0-4.0*         & NovRad19, Nes+24\\% 5823
6825 & Irvine         & 13.9 & 40    & 10  &  $1.8\pm0.5$      & PraVok09, Pra18, not listed \\% 6488
7353 & Kazuya         & 12.5 & 30    & 377 &  $\lesssim4$      & Nes+15, Car+18a \\% 6994
9332 & 1990 SB1       & 13.2 & 10    & 8   &  0.0165          & a pair and binary 2016 ER139 in Pra+19b, Nes+24\\% 8915
10164 & Akusekijima   & 12.9 & 10    & 18  &  0.5-3.0          & Nes+24\\% 9719
10321 & Rampo         & 14.4 & 10    & 47  &  $0.75\pm0.15$*    & PraVok09, Pra+18, Vok24a, Fig. \ref{case1}, not listed \\ % 9871
10484 & Hecht         & 14.0 & 10    & 3   &  $0.25\pm0.05$*    & in Vesta, Pra+19a, 2 interlopers, not listed\\% , 10032\\
11842 & Kap'bos       & 14.3 & 10    & 5   &  $<1.5$*           & PraVok09, Pra+18, Fat+20, Vok+24a, not listed  \\%11346
14627 & Emilkowalski  & 13.6 & 10    & 17  &  $0.3\pm0.1$      & double conv., NesVok06, Pra+19a, Fat+20, Vok+24a\\%14054
15156 & 2000FK38      & 13.8 & 50    & 11  &  --               & not real?, P/2006 VW139 in Nov+12\\%14573
16598 & Brugmansia    & 14.7 & 10    & 3   &  $0.17\pm0.06$    & 1992 YC2, NesVok06, Pra+18\\%15964
18429 & 1994AO1       & 13.2 & 10    & 37  &  $2.0\pm0.5$*     & Fig. \ref{case1}, NovRad19, Nes+24\\ %17728
18777 & Hobson        & 15.1 & 10    & 66  & 0.7 or 3.5*       & PraVok09, RosPla17,18, Vok+21b+24a, not listed \\%18074
20674 & 1999VT1       & 12.9 & 10    & 35  & $2.0\pm0.5$*      & nice conv., 20674 offset, P/2012, Nov+14\\ %19933
21509 & Lucascavin    & 15.0 & 10    & 3   & $\lesssim1$       & NesVok06, Pra+18, Vok+24a\\%20743
22280 & Mandragora    & 14.1 & 15    & 67  & $<2$*             & complicated conv., Pra+18, not listed\\%21481
39991 & Iochroma      & 14.6 & 10    & 7   & $<0.7$*           & double conv., PraVok09, Pra+18, not listed\\ % 38759
63440 & 2001MD30      & 15.3 & 20    & 5   & $<1$              & Pra+19a, Fat+20, not listed\\%61827
66583 & Nicandra      & 15.3 & 10    & 13  & $<3$*             & Pra+18, not listed\\ %64889 
70208 & 1999RX33      & 15.8 & 20    & 16  & 0.7-1.0*          & Nes+24\\%68462
108138 & 2001GB11     & 16.1 & 20    & 47  & $3.5\pm0.5$*      & Fig. \ref{case1}, Nes+15, Car+18a \\ %105921
157123 & 2004NW5      & 16.8 & 20    & 17  & $<2$              & some interlopers, Pra+19a, Fat+20, not listed\\ %154036 
\hline \hline\\
\end{tabular}
}
\vspace*{-8.mm}
\caption{\scriptsize Previously known young asteroid families. The ones with a star in the 6th column had their age revised
  in the present work. The third column gives the absolute magnitude of the brightest member after which the family is
  named. References:  MolHae09--Molnar \& Haegert (2009), Nes+15--Nesvorn\'y et al. (2015), Bro+24a--Bro\v{z} et al.
  (2024a), Nes+24--Nesvorn\'y et al. (2024), Nes+03--Nesvorn\'y et al. (2003), Tsi+07--Tsiganis et al. (2007),
  Car+17--Carruba et al. (2017), Bro+24b--Bro\v{z} et al. (2024b), NovRad19--Novakovi{\'c} \& Radovi{\'c} (2019),
  Vok+21a--Vokrouhlick\'y et al. (2021a), Vok+24a--Vokrouhlick\'y et al. (2024a),
  Hig+06--Higgins et al. (2006), Pra+12a,b--Pravec et al. (2012a,b), 
  Mas+13--Masiero et al. (2013), Tsi19--Tsirvoulis (2019), Carruba \& Ribeiro (2020),
  Nes+08--Nesvorn\'y et al. (2008), Car19--Carruba (2019), Nov+10--Novakovi\'c (2010),
  Nes+02--Nesvorn\'y et al. (2002), NesBot04--Nesvorn\'y \& Bottke (2004), Car16--Carruba et al. (2016),
  Nes+06c--Nesvorn\'y et al. (2006c), Vok+09--Vokrouhlick\'y et al. (2009), Vok+17--Vokrouhlick\'y et al. (2017),
  PraVok09--Pravec \& Vokrouhlick\'y (2009), Pra+19a--Pravec et al. (2019a), Vok+11--Vokrouhlick\'y et al. (2011),
  Vok+17--Vokrouhlick\'y et al. (2016), Car+18a--Carruba et al. (2018a), Pra+19b--Pravec et al. (2019b),
  Nes+03--Nesvorn\'y et al. (2003),
  Car+18b--Carruba et al. (2018b),  VokNes08--Vokrouhlick\'y \& Nesvorn\'y (2008), Nov+22--Novakovi\'c et al. (2024),
  Nov+12--Novakovi\'c et al. (2012), Vok+21b--Vokrouhlick\'y et al. (2021b),
  Pra18--Pravec et al. (2018), NesVok06--Nesvorn\'y \& Vokrouhlick\'y (2008),
  RosPla17,18--Rosaev \& Pl\'avalov\'a (2017,2018),  
  Nov+14--Novakovi\`c et al. (2012), Fat+20--Fatka et al. (2020).}
\end{table}

\clearpage

\begin{table}
\label{tab2}  
\centering
{ \footnotesize
\begin{tabular}{rlllll}
\hline \hline
Num.   & Name         & $H_{\rm 1st}$     & \# of     & $t_{\rm age}$       & Notes \\
&              & (mag)   & mem.      & (Myr)     &       \\
\hline
1346   & Gotha        & 11.5    & 7         & $0.6\pm0.1$  & interloper 149573\\ %, my id: 1310  \\
1821   & Aconcagua    & 13.3    & 5         & $0.8\pm0.2$  & \\ %1767  \\ 
5722   & Johnscherrer & 14.2    & 4         & $0.9\pm0.3$  & in Flora \\ %5425  \\
5950   & Leukippos    & 12.9    & 16        & $1.3\pm0.5$  & in Koronis, Fig. \ref{bigfam}\\ %5641  \\
5971   & Tickell      & 12.5    & 13        & $1.5\pm0.3$  & in Eunomia, Fig. \ref{bigfam}\\% 5662  \\
7629   & Foros        & 14.6    & 3         & $0.3\pm0.1$  & in Nysa-Polana\\% 7259 \\
8306   & Shoko        & 15.2    & 4         & $0.4\pm0.1$  & in Flora, pair and binary in Pra+19a\\%, 7914 \\  
14155  & Cibronen     & 14.6    & 3         & $0.5\pm0.2$   & in Flora\\%13590 \\
14161  & 1998SO145    & 15.1    & 3         & $0.7\pm0.5$   & \\%13596
14798  & 1978UW4      & 14.2    & 4         & $0.35\pm0.05$ & in Vesta\\%14220 \\
15180  & 9094P-L      & 15.2    & 3         & $0.7\pm0.2$   & \\%14597 \\
16126  & 1999 XQ86    & 13.0    & 6         & $0.15\pm0.5$  & pair in Pra+19a\\%, 15496 \\
22216  & 1242T-2      & 14.3    & 5         & $0.5\pm0.2$   & in Juno\\%21419 \\ 
22766  & 1999 AE7     & 14.1    & 5         & $<1.5$        & Nes+24, age not clear\\%, 21962 \\
23637  & 1997 AM6     & 15.6    & 3         & $<0.2$        & in Vesta, compact \\%22802 \\ 
26046  & 2104T-2      & 14.6    & 3         & $0.7\pm0.2$   & in Koronis\\%25125 \\
26170  & Kazuhiko     & 13.7    & 9         & $<2.0$        & case not clear, Nes+24\\%, 25246 \\ 
28805  & Fohring      & 14.8    & 12        & $0.7\pm0.3$  & Fig. \ref{bigfam} \\%27852 \\
28965  & 2001FF162    & 14.7    & 6         & 1.0-2.5       & in Koronis, two interlopers?\\%, 28009 \\
30301  & Kuditipudi   & 15.0    & 3         & $0.5 \pm 0.1$ & in Baptistina, pair in Pra+19a\\%, 29326 \\
34216  & 2000QK75     & 13.9    & 22        & $1.5\pm0.7$   & large\\%, 33141\\
38184  & 1999KF       & 15.3    & 4         & $0.7 \pm 0.2$ & in Euterpe, pair in Pra+19a\\%, 37025 \\
41331  & 1999XB232    & 13.6    & 18        & $1.7\pm0.5$   & in Phocaea, Fig. \ref{bigfam}, Nes+24\\%, 40082 \\
44938  & 1999VV50     & 15.3    & 3         & $<0.5$        & \\%43635
45765  & 2000LJ3      & 13.8    & 3         & 0.5-0.7       & \\%44460 \\
45974  & 2001BG35     & 15.5    & 3         & $<1$          & \\%45855   
46162  & 2001FM78     & 14.9    & 4         & 0.2-0.8       & in Maria, pair in Pra+19a, not clear\\%, 44852 \\
48939  & 1998QO8      & 15.3    & 5         & 0.1-0.5       & in Vesta\\%47595 \\
50423  & 2000DE13     & 16.0    & 4         & 0.2-0.4       & \\%49072 \\
51866  & 2001PH3      & 14.1    & 6         & $0.5 \pm 0.3$ & pair in Pra+19a\\%, 50491 \\
70512  & 1999TM103    & 15.1    & 3         & $0.7 \pm 0.3$ & in Vesta\\%68765 \\
80245  & 1999WM4      & 15.7    & 3         & $0.12\pm0.2$  & \\%78399 \\
81291  & 2000FA70     & 15.3    & 6         & $<0.6$        & complicated convergence\\%, 79441\\
86419  & 2000AL245    & 15.7    & 5         & $<1.3$        & in Hungarias\\%84483 \\
100416 & Syang        & 16.0    & 3         & $0.11\pm0.2$  & in the Hungaria fam.\\%98280 \\
100440 & 1996PJ6      & 16.4    & 3         & $0.15\pm0.10$  & in Nysa-Polana, pair in Pra+19a, compact\\%, 98304 \\
103022 & 1999XU109    & 15.2    & 7         & 2.0-4.0       & two groups in $\Omega$\\%, 100857 \\
107286 & 2001BU76     & 16.7    & 3         & $0.42\pm0.05$ & in Massalia\\%, 105076 \\
111298 & 2001XZ55     & 16.3    & 3         & $<0.2$        & in Nysa-Polana, compact \\%, 109047 \\
114555 & 2003BN44     & 15.1    & 58        & $3 \pm 1$     & in Dora, Figs. \ref{big}-\ref{big2}, 16472 \& 30693 nearby, 7 m/s\\%, 112262 \\
128637 & 2004RK22     & 15.9    & 4         & $<0.5$        & case not clear, needs yarko\\%, 126131\\
\hline \hline\\
\end{tabular}
}
\vspace*{-6.mm}
\caption{\scriptsize 63 new young asteroid families (continued in Table 3).
  These families were identified by the method described in Section 2.3.
  The membership is given here with the standard 3D HCM and $d_{\rm cutoff}=10$ m/s. Three of these families
  (22766, 26170 and 41331) were already mentioned in Nesvorn\'y et al. (2024).
  Our best age estimate for each new family is given in the fifth column. References:
  Pra+19a--Pravec et al. (2019a),  PraVok09--Pravec \& Vokrouhlick\'y (2009), Nes+24--Nesvorn\'y et al. (2024).}
\end{table}

\begin{table}
\label{tab3}  
\centering
{ \footnotesize 
\hspace*{-0.8cm}
\begin{tabular}{rlllll}
\hline \hline
Num. & Name & $H_{\rm 1st}$    & \# of   & $t_{\rm age}$   & Notes \\
     &      & (mag)   & mem.   & (Myr) &      \\
\hline
133302 & 2003SM45     & 16.4    & 3         & $0.1\pm0.1$   & in Massalia, 64480 offset in angles \\%130734\\
141906 & 2002PJ73     & 16.5    & 3         & $1.2\pm0.5$   & in Flora, 2008 UU335 offset in peri. long.\\%139128\\
156163 & 2001TF112    & 16.1    & 3         & $1.2\pm0.3$   & \\%153093 \\
180023 & 2003AU16     & 16.7    & 3         & $0.04\pm0.02$ & in Hungarias, compact\\%176447 \\
209165 & 2003UC86     & 17.0    & 4         & $<2$          & \\%204939 \\
213065 & 1999RT127    & 17.0    & 3         & $1.0\pm0.5$   & \\%208804 \\
237295 & 2008YN7      & 16.8    & 3         & $<1$          & in Clarissa, 3 asteroids nearby\\%232439 \\
237517 & 2000SP31     & 17.0    & 3         & $0.7\pm0.3$   & \\%232652 \\
238659 & 2005EG114    & 16.8    & 3         & $0.5\pm0.2$   & \\%233774 \\
267333 & 2001UZ193    & 17.3    & 3         & $<3$          & convergence unclear\\%261692 \\
267721 & 2003CL11     & 17.3    & 3         & $0.015\pm0.005$  & in the Hungaria fam., compact\\%262072 \\
346662 & 2008YG14     & 16.8    & 4         & $0.8\pm0.5$   & \\%338937
349030 & 2006VA32     & 16.3    & 3         & $0.4\pm0.2$   & \\%341250
353790 & 2012PE32     & 17.1    & 3         & $0.8\pm0.3$   & in Nysa-Polana\\%345810
%363228 & 2001WR54     & 17.2    & 4         & $<2$          & 114031 offset in nodal long. \\%354804  
368103 & 2013EN       & 17.9    & 3         & $<1$          & in the Hungaria fam. \\%359582 
381362 & 2008EP15     & 17.9    & 7         & $<2$          & 405843 brightest\\%372480
384028 & 2008UE119    & 16.7    & 3         & $<2$          & double convergence\\%375070 
481085 & 2005SX233    & 16.8    & 6         & $<1$          & compact, related to 21028?, pair in Pra+19a\\%469411
503256 & 2015KL76     & 16.6    & 14        & $<3$          & \\%490966 
572119 & 2008CF225    & 18.4    & 4         & $<1$          & \\%557712
--     & 2006SO248    & 19.1    & 5         & $<0.5$        & 496607 interloper\\%666359
--     & 2021PE78     & 18.9    & 33        & $1.1\pm0.5$   & \\%1182797
\hline \hline\\
\end{tabular}
}
\caption{Table 2 continued.}
\end{table}

\clearpage

\begin{table}
\label{tab4}  
\centering
{ \small
\begin{tabular}{rlllllll}
\hline \hline
Number & Name    & $H$      & $a_{\rm p}$ & $e_{\rm p}$ & $\sin i_{\rm p}$ & $\varpi_{\rm p}$ & $\Omega_{\rm p}$ \\
       &         & (mag)    & (au)       &           &                 &   (deg)        &  (deg)          \\
\hline
\multicolumn{8}{c}{\it 23637 1997AM6 family} \\
23637  & 1997AM6   & 15.6 & 2.30575 & 0.088782 & 0.126825 & 87.6440 & 108.933 \\
580635 & 2015CG73  & 18.8 & 2.30574 & 0.088779 & 0.126825 & 87.6440 & 108.936 \\
--     & 2015BC618 & 18.8 & 2.30573 & 0.088784 & 0.126828 & 87.6222 & 108.943 \\
\multicolumn{8}{c}{\it 100444 1996PJ6 family} \\
100444  & 1996PJ6   & 16.5 & 2.33947 & 0.200253 & 0.041162 & 17.3149 & 219.055 \\
575395  & 2011SE164 & 18.6 & 2.33945 & 0.200273 & 0.041164 & 17.2779 & 219.066 \\
--      & 2016QS128 & 18.8 & 2.33957 & 0.200214 & 0.041167 & 17.5077 & 218.947 \\
\multicolumn{8}{c}{\it 111298 2001XZ55 family} \\
111298  & 2001XZ55 & 15.9  & 2.39105 & 0.169219 & 0.041819 & 62.6433 & 125.058 \\
457548  & 2008YL20 & 18.3  & 2.39106 & 0.169194 & 0.041814 & 63.2343 & 124.897 \\
--      & 2017VL47 & 19.3  & 2.39111 & 0.169181 & 0.041815 & 63.5522 & 124.668 \\
\multicolumn{8}{c}{\it 180023 2003AU16 family} \\
180023  & 2003AU16  & 16.8 & 1.89637 & 0.039104 & 0.378883 & 13.2009 & 311.956 \\ 
--      & 2011BX142 & 19.0 & 1.89640 & 0.039110 & 0.378878 & 12.9941 & 312.063 \\
--      & 2019PG26  & 19.4 & 1.89627 & 0.039092 & 0.378889 & 13.7024 & 311.79  \\ 
\multicolumn{8}{c}{\it 267721 2003CL11 family} \\
267721  & 2003CL11  & 17.3 & 1.90499 & 0.080191 & 0.375595 & 37.9865 & 313.783 \\
--      & 2014DC113 & 19.1 & 1.90502 & 0.080232 & 0.375609 & 38.6293 & 313.317 \\
--      & 2019CV15  & 19.5 & 1.90500 & 0.080202 & 0.375596 & 38.3061 & 313.561 \\
\hline \hline\\
\end{tabular}
}
\caption{Very compact asteroid families identified in this work. Note the tight clustering in the proper
  nodal and proper perihelion longitudes. The osculating elements of members of these families
  are tightly clustered as well.
}
\end{table}     

\clearpage

\begin{table}
\label{tab5}  
\centering
{ \small
\begin{tabular}{rllllllll}
\hline \hline
Number & Name    & $H$      & $a$ & $e$ & $i$   & $\Omega$ & $\omega$ & $M$\\
       &         & (mag)    & (au)&     & (deg) &   (deg)  &  (deg)   & (deg) \\
\hline
\multicolumn{9}{c}{\it 25435 1999 WX3 family} \\
25435 & 1999 WX3   & 15.17  & 2.298011 & 0.186480 & 4.97120 & 211.5154 & 212.1530 & 114.4559 \\
--    & 2009 SD429 & 19.52  & 2.298113 & 0.186494 & 4.97120 & 211.5234 & 212.1462 & 128.3220 \\
--    & 2019 SY257 & 21.34  & 2.298311 & 0.186291 & 4.97217 & 211.5956 & 212.0470 & 173.6913 \\
\hline \hline\\
\end{tabular}
}
\caption{The osculating elements of three members of the 25435 1999 WX3 family (MJD 60800). The
  nominal orbit is listed for 2019 SY257. This object has a relatively large orbital
  uncertainty and not all digits given here are significant (Section 3.5).}
\end{table}     

\clearpage
\clearpage

\clearpage

\clearpage

\clearpage

\clearpage

\clearpage

%\clearpage
%\begin{figure}
%\epsscale{0.6}
%\plotone{Figs/5950.eps}
%\plotone{Figs/5971.eps}
%\caption{5950 Leukippos and 5971 Tickel.}
%\label{bigfam2}
%\end{figure}

%\clearpage
%\begin{figure}
%\epsscale{0.24}
%\plotone{5722_Johnscherrer.eps}
%\plotone{7629_Foros.eps}
%\plotone{8306_Shoko.eps}
%\plotone{10484_Hecht.eps}
%\plotone{Figs/conv_5722.eps}
%\plotone{Figs/conv_7629.eps}
%\plotone{Figs/conv_8306.eps}
%\plotone{Figs/conv_10484.eps}
%\plotone{Figs/lines_5722.eps}
%\plotone{Figs/lines_7629.eps}
%\plotone{Figs/lines_8306.eps}
%\plotone{Figs/lines_10484.eps}
%\plotone{clone_5722.eps}
%\plotone{clone_7629.eps}
%\plotone{clone_8306.eps}
%\plotone{clone_10484.eps}
%\caption{Age estimated from the past convergence of proper angles.}
%\label{prop1}
%\end{figure}

%\clearpage
%\begin{figure}
%\epsscale{0.24}
%\plotone{Figs/obr1.JPG}
%\plotone{Figs/obr2.JPG}
%\plotone{Figs/obr3.JPG}
%\plotone{Figs/obr4.JPG}
%\plotone{Figs/obr5.JPG}
%\plotone{Figs/obr6.JPG}
%\plotone{Figs/obr7.JPG}
%\plotone{Figs/obr8.JPG}
%\plotone{Figs/obr9.JPG}
%\plotone{Figs/obr10.JPG}
%\plotone{Figs/obr11.JPG}
%\plotone{Figs/obr12.JPG}
%\plotone{Figs/obr13.JPG}
%\plotone{Figs/obr14.JPG}
%\plotone{Figs/obr15.JPG}
%\plotone{Figs/obr16.JPG}
%\caption{Age estimated from the past convergence of proper angles.}
%\label{prop1}
%\end{figure}

\clearpage

%\clearpage
%\begin{figure}
%\epsscale{0.49}
%\plotone{Fig2/7629.eps}
%\plotone{Fig2/14798.eps}
%\plotone{Fig2/15180.eps}
%\plotone{Fig2/16126.eps}
%\plotone{Fig2/30301.eps}
%\plotone{Fig2/45765.eps}
%\caption{}
%\label{clone123}
%\end{figure}

%\clearpage
%\begin{figure}
%\epsscale{0.49}
%\plotone{Fig2/70512.eps}
%\plotone{Fig2/80245.eps}
%\plotone{Fig2/81291.eps}
%\plotone{Fig2/100416.eps}
%\plotone{Fig2/107286.eps}
%\plotone{Fig2/141906.eps}
%\caption{}
%\label{clone112}
%\end{figure}

%\begin{figure}
%\epsscale{0.49}
%%\plotone{Figs/5478.eps}
%%\plotone{Figs/5722.eps}
%%\plotone{Figs/7629.eps}
%%\plotone{Figs/8306.eps}
%%\plotone{Figs/10484.eps}
%\plotone{Figs/7629.eps}
%\plotone{Figs/14798.eps}
%\plotone{Figs/15180.eps}
%\plotone{Figs/16126.eps}
%%\plotone{Figs/23637.eps}
%\plotone{Figs/30301.eps}
%\plotone{Figs/45765.eps}
%%\plotone{Figs/46162.eps}
%%\plotone{Figs/111298.eps}
%\caption{Tests of convergence for new young families.}
%\label{davok}
%\end{figure}

%\begin{figure}
%\epsscale{0.49}
%%\plotone{Figs/15180.eps}
%%\plotone{Figs/26046.eps}
%\plotone{Figs/70512.eps}
%\plotone{Figs/80245.eps}
%\plotone{Figs/81291.eps}
%\plotone{Figs/100416.eps}
%\plotone{Figs/107286.eps}
%\plotone{Figs/141906.eps}
%\caption{Tests of convergence for new young families.}
%\label{davok2}
%\end{figure}

\clearpage

\clearpage

\clearpage

\clearpage

\clearpage

\end{document}